\documentclass[conference,letterpaper]{IEEEtran}

\usepackage{fancyhdr}

\usepackage{booktabs}   %% For formal tables:
\usepackage{subcaption} %% For complex figures with subfigures/subcaptions

\usepackage{amsfonts,proof,qtree,amsmath,mathtools}
\usepackage{amsthm}
\usepackage{cite}
\usepackage{flushend}
\usepackage{amssymb}
\usepackage{latexsym}
\usepackage{graphicx}
\usepackage{listings}
\usepackage{float}
\usepackage{multirow}
\usepackage[scaled]{helvet}
\usepackage[noend]{algorithmic}
\usepackage{mathrsfs}
\usepackage{mathpartir}
\usepackage{dsfont} 
\usepackage{stmaryrd}
\usepackage{url}
\usepackage{textcomp} 
\usepackage{caption}
\usepackage{multicol}
\usepackage{blindtext}
\usepackage{alltt} 
\usepackage{bbm}
\usepackage{alltt}
\usepackage{verbdef}
\usepackage{xspace}
\usepackage{verbatim}
\usepackage{enumitem}
\usepackage{lipsum}
\usepackage{wrapfig}
\usepackage{xcolor}
\usepackage{hyperref}
\hypersetup{linkcolor=black,citecolor=black,urlcolor=RubineRed}

\usepackage{enumitem}
\usepackage{pifont}
\usepackage[scaled]{helvet}
\usepackage{varwidth}
\definecolor[named]{ACMBlue}{cmyk}{1,0.1,0,0.1}
\definecolor[named]{ACMYellow}{cmyk}{0,0.16,1,0}
\definecolor[named]{ACMOrange}{cmyk}{0,0.42,1,0.01}
\definecolor[named]{ACMRed}{cmyk}{0,0.90,0.86,0}
\definecolor[named]{ACMLightBlue}{cmyk}{0.49,0.01,0,0}
\definecolor[named]{ACMGreen}{cmyk}{0.20,0,1,0.19}
\definecolor[named]{ACMPurple}{cmyk}{0.55,1,0,0.15}
\definecolor[named]{ACMDarkBlue}{cmyk}{1,0.58,0,0.21}

\hypersetup{colorlinks,
  linkcolor=ACMDarkBlue,
  citecolor=ACMPurple,
  urlcolor=ACMDarkBlue,
  filecolor=ACMDarkBlue}

\newtheorem{definition}{Definition}

\setlist[itemize]{leftmargin=*}

\begin{document}

\setlength{\parindent}{0.15in}
\setlength{\topsep}{0cm}
\setlength{\parskip}{0pt}

\setlist[enumerate]{topsep=0pt,itemsep=-1ex,partopsep=1ex,parsep=1ex,itemindent=0pt,leftmargin=10pt}
\setlist[itemize]{topsep=0pt,itemsep=-1ex,partopsep=1ex,parsep=1ex,itemindent=0pt,leftmargin=8pt}

\newcommand{\authnote}[2]{{\bf \textcolor{blue}{#1}: \em \textcolor{red}{#2}}}
\newcommand{\subtbl}[1]{{\begin{tabular}[c]{@{}l@{}}#1\end{tabular}}}
\newcommand{\yes}[0]{{$\checkmark$}}
\newcommand{\no}[0]{{$\times$}}
\newcommand{\lin}[0]{linearizability\xspace}
\newcommand{\Lin}[0]{Linearizability\xspace}
\newcommand{\syn}[0]{synchronization\xspace}
\newcommand{\Syn}[0]{Synchronization\xspace}

%\newcolumntype{L}{>{\centering\arraybackslash}m{18mm}}

% \makeatletter
% \patchcmd{\@makecaption}
%   {\scshape}
%   {}
%   {}
%   {}
% \makeatother

%% Uncomment the following line to remove author notes from the paper
%% for submission
%% \renewcommand{\authnote}[2]{}

\renewcommand{\paragraph}[1]{\par\noindent\textbf{#1}.}
\newcommand{\cmark}{\ding{51}}%
\newcommand{\xmark}{\ding{55}}%
\newcommand{\tab}[1]{\hspace{.2\textwidth}\rlap{#1}}
\newcommand{\itab}[1]{\hspace{0em}\rlap{#1}}

\newcommand{\huhong}[1]{\authnote{huhong}{#1}}
\newcommand{\zhenkai}[1]{\authnote{zhenkai}{#1}}
\newcommand{\prateek}[1]{\authnote{prateek}{#1}}
\newcommand{\aashish}[1]{\textcolor{green}{#1}}
\newcommand{\note}[1]{\textcolor{blue}{#1}}

\newcommand{\squish}{
\setlength{\topsep}{0pt}
\setlength{\itemsep}{0ex}
\setlength{\parskip}{0pt}
}

\renewcommand{\lstlistingname}{Code}
\newcommand{\doplang}{{\sc MinDOP}\xspace}

\newtheorem{myrule}{Rule}
\hyphenation{op-tical net-works semi-conduc-tor}

\definecolor{lightgray}{rgb}{.95,.95,.95}
\definecolor{darkgray}{rgb}{.4,.4,.4}
\definecolor{purple}{rgb}{0.65, 0.12, 0.82}
\definecolor{mygreen}{rgb}{0,0.6,0}

\lstdefinelanguage{Assembly}{
    keywords={MOVE, STORE, ADD, and, brk, read, write, cmp, mov, add, ret, false, for, if, else, in, int, lea, then, continue, dec, jnz, while, char, printf, void},
    keywordstyle=\color{blue}\bfseries,
    ndkeywords={assert, condition, class, export, boolean, throw, implements, import, this},
    ndkeywordstyle=\color{red}\bfseries,
    identifierstyle=\color{black},
    sensitive=false,
    comment=[l]{//},
    morecomment=[s]{/*}{*/},
    commentstyle=\color{purple}\ttfamily,
    stringstyle=\color{red}\ttfamily,
    morestring=[b]',
    morestring=[b]"
}

\lstset{
   language=C,
   backgroundcolor=\color{lightgray},
   keywordstyle=\color{blue}\bfseries,
   extendedchars=true,
   basicstyle=\footnotesize\ttfamily,
   frame=single,
   framexleftmargin=-4pt,
   framexrightmargin=-4pt,
   showstringspaces=false,
   showspaces=false,
   commentstyle=\color{purple}\bfseries,
   numbers=left,
   numberstyle=\footnotesize,
   numbersep=-10pt,
   tabsize=2,
   breaklines=true,
   showtabs=false,
   captionpos=b
}

%\newcommand\mycommfont[1]{\footnotesize\textcolor{blue}{#1}}
%\SetAlFnt{\small}
%\SetCommentSty{mycommfont}
%\DontPrintSemicolon
%\renewcommand{\thealgocf}{}

\hyphenation{Dist-Algo}
\hyphenation{Ssref-lect}

% Colors

\definecolor{shadecolor}{gray}{1.00}
\definecolor{ddarkgray}{gray}{0.75}
\definecolor{darkgray}{gray}{0.30}
\definecolor{light-gray}{gray}{0.87}
\newcommand{\whitebox}[1]{\colorbox{white}{#1}}
\newcommand{\graybox}[1]{\colorbox{light-gray}{#1}}
\newcommand{\darkgraybox}[1]{\colorbox{ddarkgray}{#1}}
\newcommand{\gbm}[1]{\graybox{${#1}$}}
\newcommand{\Inv}{\text{\textsc{Inv}}}
\newcommand{\TPCInv}{\text{\textsc{TPCInv}}}

\newcommand{\fstar}{\text{F}^{\star}}
\newcommand{\etc}{\emph{etc}}
\newcommand{\ie}{\emph{i.e.}\xspace}
\newcommand{\Ie}{\emph{I.e.}\xspace}
\newcommand{\eg}{\emph{e.g.}\xspace}
\newcommand{\Eg}{\emph{E.g.}\xspace}
\newcommand{\vs}{\emph{vs.}\xspace}
\newcommand{\etal}{\emph{et~al.}\xspace}
\newcommand{\adhoc}{\emph{ad hoc}\xspace}
\newcommand{\viz}{\emph{viz.}\xspace}
\newcommand{\dom}[1]{\mathsf{dom}(#1)}
\newcommand{\aka}{\textit{a.k.a.}\xspace}
\newcommand{\cf}{\textit{cf.}\xspace}
\newcommand{\wrt}{\emph{wrt.}\xspace}
\newcommand{\Iff}{\emph{iff}\xspace}
\newcommand{\loef}{L\"{o}f}
\newcommand{\edeff}{\triangleq}

\newcommand{\denot}[1]{\llbracket{#1}\rrbracket}

\newcommand{\ifext}[2]{\ifdefined\extflag{#1}\else{#2}\fi}
\newcommand{\ifcomm}[1]{\ifdefined\extcomm{#1}\else{}\fi}

\theoremstyle{plain}
\newenvironment{proofsketch}{\trivlist\item[]\emph{Proof sketch}:}%
{\unskip\nobreak\hskip 1em plus 1fil\nobreak$\square$
\parfillskip=0pt%
\endtrivlist}

\newcommand{\mute}[1]{{#1}}

% remarks
\newcommand{\todo}[1]{\mute{\textcolor{red}{}}}
\newcommand{\ak}[1]{\mute{\textcolor{ACMGreen}{(Aashish: {#1})}}}
\newcommand{\inik}[1]{\mute{\textcolor{ACMPurple}{(Ivica: {#1})}}}
\newcommand{\ps}[1]{\mute{\textcolor{ACMGreen}{(Prateek: {#1})}}}
\newcommand{\ah}[1]{\mute{\textcolor{ACMLightBlue}{(Aquinas: {#1})}}}
\newcommand{\is}[1]{\mute{\textcolor{ACMBlue}{(Ilya: {#1})}}}

\newcommand{\tsend}{\mathsf{send}}
\newcommand{\treceive}{\mathsf{receive}}
\newcommand{\tskip}{\mathsf{skip}}

\newcommand{\disel}{{\sc Disel}\xspace}
\newcommand{\withinvr}{{\sc WithInv}\xspace}
\newcommand{\from}{\mathsf{from}}
\newcommand{\tto}{\mathsf{to}}
\newcommand{\True}{\mathsf{True}}
\newcommand{\id}{\mathsf{id}}
\newcommand{\False}{\mathsf{False}}
\newcommand{\set}[1]{\left\{{#1}\right\}}
\newcommand{\angled}[1]{\langle{#1}\rangle}

%% Specs
\newcommand{\specK}[1]{\ensuremath{\textcolor{blue}{\set{#1}}}}
\newcommand{\pre}[2]{\specK{\lambda {#1}.~{#2}}}
\newcommand{\post}[3]{\specK{\lambda {#1}~{#2}.~{#3}}}
\newcommand{\psep}{\ast}

%% Protocols
\newcommand{\ppr}{\mathcal{P}}
\newcommand{\cpr}{\mathcal{C}}

%% State-space and Transitions
\newcommand{\treq}{\mathsf{Req}}
\newcommand{\tresp}{\mathsf{Resp}}
\newcommand{\acts}{\Send}
\newcommand{\actr}{\Recv}
\newcommand{\trans}[1]{\mathit{#1}}
\newcommand{\To}{\mathsf{to}}
\newcommand{\Rs}{\mathsf{rs}}
\newcommand{\From}{\mathsf{from}}
\newcommand{\args}{\mathit{args}}
\newcommand{\ans}{\mathit{ans}}
\newcommand{\aand}{\wedge}
\newcommand{\aandb}{\&\!\!\!\&}
\newcommand{\oor}{\vee}
\newcommand{\body}{\mathit{body}}
\newcommand{\clients}{\overline{C}}
\newcommand{\servers}{\overline{S}}
\newcommand{\Send}{\mathsf{send}}
\newcommand{\Recv}{\mathsf{recv}}
\newcommand{\csend}[1]{\Send[#1]}
\newcommand{\creceive}[1]{\Recn[#1]}
\newcommand{\lab}{\ell}
\newcommand{\Some}{\mathsf{Some}}
\newcommand{\None}{\mathsf{None}}
\newcommand{\Truez}{\mathsf{True}}
\newcommand{\Falsez}{\mathsf{False}}

\newcommand{\fun}{\rightarrow}
\newcommand{\join}{\uplus}

% Semantics
\newcommand{\nstep}[2]{\mathrel{\overset{#2}{~\leadsto_{#1}~}}}
\newcommand{\nsem}[4]{{#3}\mathrel{\overset{#2}{\leadsto_{#1}}}{#4}}
\newcommand{\ninf}[4]{{#3}\mathrel{\overset{\!\!\!\neg{#2}*}{\leadsto_{#1}}}{#4}}
\newcommand{\nsemw}[4]{{#3}\mathrel{\overset{w, #2}{\leadsto_{#1}}}{#4}}

% Java listing
\definecolor{pblue}{rgb}{0.13,0.13,1}
\definecolor{pgreen}{rgb}{0,0.5,0}
\definecolor{pred}{rgb}{0.9,0,0}
\definecolor{pgrey}{rgb}{0.46,0.45,0.48}

\definecolor{ckeyword}{HTML}{7F0055}
\definecolor{ccomment}{HTML}{3F7F5F}
\definecolor{cnumber}{HTML}{2A0099}

\definecolor{verylightgray}{rgb}{.97,.97,.97}

\lstdefinelanguage{Solidity}{
	keywords=[1]{anonymous, synchronized, assembly, assert, break, call, callcode, case, catch, class, constant, continue, contract, debugger, default, delegatecall, delete, do, else, event, export, external, finally, for, function, gas, if, implements, import, in, indexed, instanceof, interface, internal, is, length, library, log0, log1, log2, log3, log4, memory, modifier, new, payable, pragma, private, protected, public, pure, push, require, return, returns, revert, selfdestruct, send, storage, struct, suicide, super, switch, then, throw, transfer, try, typeof, using, view, while, with, addmod, ecrecover, keccak256, mulmod, ripemd160, sha256, sha3}, % generic keywords including crypto operations
	keywordstyle=[1]\color{blue}\bfseries,
	keywords=[2]{address, bool, byte, bytes, bytes1, bytes2, bytes3, bytes4, bytes5, bytes6, bytes7, bytes8, bytes9, bytes10, bytes11, bytes12, bytes13, bytes14, bytes15, bytes16, bytes17, bytes18, bytes19, bytes20, bytes21, bytes22, bytes23, bytes24, bytes25, bytes26, bytes27, bytes28, bytes29, bytes30, bytes31, bytes32, enum, int, int8, int16, int24, int32, int40, int48, int56, int64, int72, int80, int88, int96, int104, int112, int120, int128, int136, int144, int152, int160, int168, int176, int184, int192, int200, int208, int216, int224, int232, int240, int248, int256, mapping, string, uint, uint8, uint16, uint24, uint32, uint40, uint48, uint56, uint64, uint72, uint80, uint88, uint96, uint104, uint112, uint120, uint128, uint136, uint144, uint152, uint160, uint168, uint176, uint184, uint192, uint200, uint208, uint216, uint224, uint232, uint240, uint248, uint256, var, void, ether, finney, szabo, wei, days, hours, minutes, seconds, weeks, years},	% types; money and time units
	keywordstyle=[2]\color{teal}\bfseries,
	keywords=[3]{block, blockhash, coinbase, difficulty, gaslimit,
    number, timestamp, this, true, false, msg, data, gas, sender, value, sig, value, balance, now, tx, gasprice, origin},	% environment variables
	keywordstyle=[3]\color{violet}\bfseries,
	identifierstyle=\color{black},
	sensitive=false,
	comment=[l]{//},
	morecomment=[s]{/*}{*/},
	commentstyle=\color{gray}\ttfamily,
	stringstyle=\color{red}\ttfamily,
	morestring=[b]',
	morestring=[b]"
}

\lstset{
	language=Solidity,
	backgroundcolor=\color{white},
	extendedchars=true,
	basicstyle=\scriptsize\sffamily,
	%basicstyle=\fontsize{6}{9}\ttfamily,
  escapeinside={\%*}{*)},
	showspaces=false,
	numbers=left,
	numberstyle=\sffamily\tiny,
	numbersep=7pt,
	tabsize=2,
	breaklines=true,
	showtabs=false,
	captionpos=b,
	xleftmargin=5em,
	frame=none,
	framexleftmargin=0.2em,
	escapechar=\$
	% postbreak=\mbox{\textcolor{red}{$\hookrightarrow$}\space\space\space\space},
}
\newcommand{\scode}[1]{\lstinline[language=Solidity,basicstyle=\small\ttfamily]{#1}}
\newcommand{\fcode}[1]{\lstinline[language=Solidity,basicstyle=\footnotesize\ttfamily]{#1}}
\newcommand{\code}[1]{\scode{#1}}
\newcommand{\mcode}[1]{\lstinline[language=Solidity,basicstyle=\small\ttfamily,mathescape]{#1}}

\newcommand{\ihb}[1]{\mathit{ihb}(#1)}
\newcommand{\ihbt}[2]{\mathit{ihb}_{#1}(#2)}
\newcommand{\HB}{\emph{hb}\xspace}

\newcommand{\IHB}{\emph{ihb}\xspace}
\newcommand{\Me}{\mathit{xcl}}
\newcommand{\me}[1]{\Me(#1)}
\newcommand{\ME}{\emph{xcl}\xspace}
\newcommand{\ctx}{\mathit{ctx}}
\newcommand{\fn}{\mathit{fn}}
\newcommand{\fname}{\mathsf{fname}}
\newcommand{\enabled}[1]{\mathit{enabled}({#1})}

\newcommand{\estep}[1]{\xrightarrow[\phantom{aa}]{#1}}
\newcommand{\tstep}[1]{\xRightarrow[\phantom{aa}]{#1}}
\newcommand{\exn}{\lightning}

\newcommand{\trace}{\tau}
\newcommand{\rulename}[1]{\textsc{#1}}
\newcommand{\tname}[1]{\textsc{#1}\xspace}
\newcommand{\pname}[1]{{\small{\textsf{#1}}}\xspace}
\newcommand{\oyente}{\tname{Oyente}}

\newcommand{\codename}{\toolname}
\newcommand{\toolname}{\tname{EthRacer}}
\newcommand{\tool}{\toolname}
%\newcommand{\toolname}{\tname{Corriente}}
% Why not?

\newcommand{\Oraclize}{\textsf{Oraclize}\xspace}
\newcommand{\oraclize}{linearizability optimizer\xspace}
\newcommand{\erc}{{ERC-20}\xspace}

\newcommand{\hbord}{\xrightarrow[]{\text{hb}}}
\newcommand{\Hb}{\mathit{hb}}
\newcommand{\hb}[2]{\Hb({#1},{#2})}
\newcommand{\whb}[2]{\mathit{whb}({#1},{#2})}

\newcommand{\elite}{\tname{EtherLite}}
\newcommand{\Id}{\mathit{id}}
\newcommand{\pc}{\mathsf{pc}}
\newcommand{\fld}{\mathsf{fld}}
\newcommand{\bstate}{\sigma}
\newcommand{\conf}{\delta}
\newcommand{\Bal}{\mathsf{bal}}
\newcommand{\Code}{\mathsf{code}}
\newcommand{\Sender}{\mathsf{sender}}
\newcommand{\Value}{\mathsf{value}}
\newcommand{\String}{\mathit{string}}
\newcommand{\Data}{\mathsf{data}}
\newcommand{\Nat}{\mathbb{N}}
\newcommand{\many}[1]{\overline{#1}}
\newcommand{\cstate}{\rho}
\newcommand{\opn}[1]{{\tt {#1}}}
\newcommand{\Call}[1]{\mathsf{call}({#1})}
\newcommand{\stepc}{\Longrightarrow}
\newcommand{\Lab}{\ell}
\newcommand{\clos}[1]{{#1}^{*}}
\newcommand{\ab}[1]{\widehat{#1}}
\newcommand{\Index}{\mathtt{index}}

%% Title information
\title{Exploiting The Laws of Order in Smart Contracts}

%% [Short Title] is optional; when present, will be used in header
%% instead of Full Title.
% \titlenote{with title note}             %% \titlenote is optional;
%                                         %% can be repeated if necessary;
%                                         %% contents suppressed with 'anonymous'
% \subtitle{Subtitle}                     %% \subtitle is optional
% \subtitlenote{with subtitle note}       %% \subtitlenote is optional;
%                                         %% can be repeated if necessary;
%                                         %% contents suppressed with 'anonymous'

\author{ \IEEEauthorblockN {Aashish Kolluri \IEEEauthorrefmark{1}, Ivica Nikoli\'c \IEEEauthorrefmark{1}, Ilya Sergey \IEEEauthorrefmark{2}, Aquinas Hobor \IEEEauthorrefmark{1} \IEEEauthorrefmark{3}, Prateek Saxena \IEEEauthorrefmark{1}}\
\IEEEauthorblockA{ \IEEEauthorrefmark{1}School Of Computing, NUS, Singapore\\ \IEEEauthorrefmark{3}Yale-NUS College, Singapore\\ \IEEEauthorrefmark{2}University College London, United Kingdom}
}

\maketitle

\begin{abstract}
We investigate a family of bugs in
blockchain-based smart contracts, which we call \emph{event-ordering} (or EO)
bugs. These bugs are intimately related to the dynamic ordering
of contract \emph{events}, \ie, calls of its functions on the blockchain, and
enable potential exploits of millions of USD worth of Ether. Known
examples of such bugs and prior techniques to detect them have been restricted to a
small number of event orderings, typicall 1 or 2. Our work provides a new formulation 
of this general class of EO bugs as finding concurrency properties 
arising in {\em long} permutations of such events. 

The technical challenge in detecting our formulation of EO bugs is the
inherent combinatorial blowup in path and state space analysis, even for simple
contracts. We propose the first use of partial-order reduction techniques,
using happen-before relations extracted automatically for contracts, along
with several other optimizations built on a dynamic symbolic execution technique.
We build an automatic tool called \toolname that requires no hints from users and runs directly on Ethereum
bytecode. It flags 7-11\% of over {\em ten thousand}  contracts analyzed
in roughly 18.5 minutes per contract, providing compact event traces that human analysts
can run as witnesses. These witnesses are so compact that confirmations require only a few
minutes of human effort. Half of the flagged contracts have subtle EO bugs, including
in ERC-20 contracts that carry hundreds of millions of dollars worth of Ether.
Thus, \toolname is effective at detecting a subtle yet dangerous class
of bugs which existing tools miss.

% Following the \emph{contracts-as-concurrent-objects} analogy, we
% examine how two key ideas originating from concurrent programming,
% linearizability and happens-before orderings can uncover EO bugs.

% We provide a formal model for capturing EO bugs.  We use our model to
% identify contracts that lack linearizability, \emph{i.e.} where the
% programmer's assumption that a pair of events always execute
% back-to-back without an intervening event is violated.
% %
% Further, we describe an algorithm for
% inferring the happens-before relation of an arbitrary contract.
% %
% We implement our insights in \toolname, a tool for dynamic analysis of
% Ethereum smart contracts for potential EO bugs.
% %
% \toolname 
%
%harden contracts against .

\end{abstract}

\section{Introduction}
\label{sec:introduction}

% 1. Intro to blockchain, cryptocurrencies and contracts.
A blockchain/cryptocurrency protocol enables a distributed network of mutually-untrusting
computational nodes (\emph{miners}) to agree on the current state and complete
history of a replicated public ledger.  The dominant consensus algorithm
was invented to facilitate decentralized payments in virtual
currencies~\cite{Nakamoto:08}, but it has since been extended to
the decentralized applications commonly known as
\emph{smart contracts}~\cite{Szabo:96}.
A typical smart contract on a blockchain is a stateful program, \ie, a
package of code and the mutable data that describes the contract's
current state, similar to an object in an OOP language.  Both the code
and the data are stored in a replicated fashion on the blockchain.
Every smart contract transaction (invocation of contract code) is
totally ordered, as agreed upon by a majority of miners, and
replicated across the system.
%

% 2. More specifically - what are Ethereum smart contracts
Smart contracts implement some domain-specific logic to act as
automatic and trustworthy mediators.  Typical applications include
multi-party accounting, voting, arbitration mechanisms, auctions, and
puzzle-solving games with distribution of rewards.
The dominant smart contract-enabled blockchain today is
Ethereum~\cite{Gavin-al:yellow-paper}, whose native token Ether has a
market capitalization over $20.5$ billion USD.\footnote{As of this
  writing, one Ether is 203 USD.}  Over a million smart
contracts have been deployed to Ethereum's blockchain.
%Due to a large number of high-value targets, n
Numerous publicly reported attacks have resulted in hundreds of
millions dollars' worth of Ether being stolen or otherwise
lost~\cite{theDao,Luu-al:CCS16}. Further, contracts cannot be patched
once deployed. This emphasizes the importance of pre-deployment
security audit and analysis of smart contracts.
%
% 3. What is the problem we're looking at?

This paper investigates a class of vulnerabilities in smart contracts
that arise due to their inherent {\em concurrent} execution model.
%Ethereum smart contracts have a complex execution model.  
Contracts can be invoked by multiple users concurrently, and the
ordering of multiple submitted transactions is non-deterministically
decided by miners through a consensus protocol.
Contracts can invoke other contracts {\em synchronously} and call
off-chain services {\em asynchronously} which return in no
pre-determined order. As Ethereum contracts are {\em stateful},
mutations of contract data persist between invocations.  Therefore,
predicting the result from a set of transactions invoking a contract
requires reasoning about the non-deterministic order of
concurrently-interacting transactions.  Developers often write
contracts assuming a certain serialized execution order of contracts,
missing undesirable behaviors only observable in complex
interleavings. Such reasoning has classically been difficult for human
auditors.  Accordingly, tools that allow developers, auditors, and
smart contract users to increase confidence that contracts behave as
expected are useful.

Certain concurrency bugs in Ethereum smart contract are known.  For
instance, prior work has highlighted how a pair of transactions, when
reordered, can cause contracts to exhibit differing Ether transfers as
output~\cite{Luu-al:CCS16}. Similarly, susceptibility of contracts to
asynchronous callbacks has been identified
previously~\cite{Bansal-al:TACAS18,Sergey-Hobor:WTSC17}.  However, the
full generality of bugs arising from unexpected ordering of {\em
  events} --- \ie calls to contract functions invoked via transactions
and callbacks --- has neither been systematically tested nor fully
understood yet.

The key encumbering challenge is that analyzing contracts under
multiple events spread over many transactions leads to {\em
  combinatorial blowup} in state-space to be checked. Existing tools
are thus designed to avoid search of large path space, by checking for
properties of often single or a pair of events. For instance, the
infamous re-entrancy bugs such as \texttt{theDao} can be found with
checking if a function can call itself within a single transaction
execution~\cite{Grossman-al:POPL18}, while transaction ordering bugs
reported by the \oyente tool check a pair of
events~\cite{Luu-al:CCS16}.
% Below phrase is not true, Maian checks arbitrary number of events
%; and similarly, safety/liveness bugs
%recently highlighted are checkable with two events of a specific type
%and order~\cite{Nikolic-al:Maian}.

\paragraph{Problem \& Approach}
In this work, we develop new and efficient analysis techniques for
Ethereum smart contracts under \emph{multiple} events. Our work
generalizes beyond several previous classes or errors into a broader
category of concurrency errors we call \emph{event-ordering} (EO)
bugs.  The core idea is to check whether changing the ordering of
input events (function invocations) of a contract results in
\emph{differing} outputs. If a contract exhibits differing outputs
under reordered input events, 
%the contract may has subtle unintended
%behavior 
it is flagged as an EO bug; otherwise, event re-ordering
produces no impact on outputs and so the contract is EO-safe.

Our formulation of EO bugs deepens the connection between contracts
and concurrent objects in traditional programming
languages~\cite{Sergey-Hobor:WTSC17,Grossman-al:POPL18}.  Specifically,
we show how to directly juxtapose the events of a contract with atomic
operations of shared memory objects.  This results in phrasing
properties of contracts directly as traditional concurrency
properties.  For instance, we observe that asynchronous calls to
off-chain services follow a \emph{call/return} pattern and can lead to
a violation of
\emph{\lin}~\cite{Herlihy-Wing:TOPLAS90,Shacham-al:OOPSLA11}.
Further, we show that contracts are susceptible to event races because
of improper enforcement of \emph{\syn}
properties~\cite{Raychev-al:OOPSLA13,Dimitrov-al:PLDI14}.  We find
hundreds of live contracts with previously unknown errors,
highlighting that programmers often desire these properties but fail
to enforce them in implementation.

To tackle combinatorial path and state explosion, we develop a number
of optimization techniques. Furthering the
``contracts-as-concurrent-objects'' analogy, we show that
partial-order reduction techniques can be applied to contract
analysis. Specifically, if two functions can only be invoked in
certain order (or else an exception results), or re-ordering a set of
functions yields the same output, then event combinations eliminate
repeatedly enumerating such sequences. This concept is captured by the
classical \emph{happens-before} (HB)
relation~\cite{Lamport:CACM78}. Unlike traditional programming
languages which have explicit synchronization primitives, 
smart contracts try to implement desirable concurrency controls using
ad-hoc program logic and global state. We show how to recover the
intrinsic HB-relation encoded in contract logic, and that it
substantially reduces the event combinations to check.

\paragraph{\codename}
Our central practical contribution is an automatic tool to find EO
bugs called \codename. We use it to measure the prevalence of these
vulnerabilities over ten thousand contracts.
Less than $1\%$ of live contracts are accompanied by source code;
hence, our tool does not require source and analyzes Ethereum bytecode
directly. This enables third-party audit and testing of contracts
without source.
We take a dynamic testing approach, constructing inputs and systematically
trying all possible function orderings until a budgeted timeout is reached.
Done na\"{i}vely, this approach would quickly lead to an intractable analysis
even for relatively small contracts, since $N$ function calls to a contract
can have $N!$ orderings.  Our approach combines symbolic execution of contract
events with fast randomized fuzzing of event sequences. Our key optimizations
exponentially reduce the search space by eliminating orderings which violate
the recovered HB-relation between events or re-order pure (side-effect-free)
events.  Further, \tool prioritizes the search for \lin violations in
asynchronous callbacks.
\codename reports only true EO violations, accompanied by
witnesses of event values that can be concretely executed. 

%exhibiting differing outputs.

%  Each executed path is analyzed using symbolic
% execution to avoid program state-space search.
% %
% %
% We devise a procedure to automatically extract the inherent
% ``happens-before'' (HB) relation encoded in contracts. Our perspective
% is that according to a contract implementor's insight, some pairs of
% events $(e_1, e_2)$ are coded to happen only in a certain order, which
% rcorresponds to an HB-ordering, \eg, $\hb{e_1}{e_2}$.

\paragraph{Empirical Results}
First, we show that most contracts {\em do not} exhibit differences in
outputs, and that when contracts {\em do} exhibit different outputs
upon re-ordering, they are likely to have an unintended behavior in
more than $50\%$ of cases. We find a total of $789$ ($7.89\%$)
violations of \syn properties and $47$ ($11\%$) violations of \lin
properties in our analysis of over ten thousand contracts. Therefore,
our formulation of properties catches a subtle and dangerous class of
EO bugs, without excessively triggering alarms.

Second, we show that our characterization of EO bugs substantially
generalizes beyond bugs known from prior works. A direct comparison to
prior work on \oyente, which focuses on specific sub-class of \syn
violations, shows that \codename find \emph{all the} $78$ true EO bugs
Oyente finds and many more ($674$ in total) which are not detected by
prior work.  Similarly, it finds $47$ \lin violations that are not
captured by any prior work, generalizing this class of errors beyond
DAO-style re-entrancy~\cite{theDao} or id-tracking bugs that fail to
pair asynchronous calls and callbacks~\cite{Sergey-Hobor:WTSC17}. We
find many bugs in popular \erc~\cite{ERC20} compliant contract
(ubiquitously used for ``ICO''s), casino/lottery, bounties, contests,
and escrow service contracts. The flagged contracts include both old
and recent; these contracts have processed millions of transactions,
holding hundreds of millions of dollars worth of Ether over their
lifetime. The results stem directly from our efficiency-enhancing
techniques --- for $95\%$ of the contracts we analyzed, \codename
produced results at an average of $18.5$ minutes per contract.

Lastly, our \codename minimizes human effort in analysis of results.
When \codename reports EO violations, it provides concrete witnesses
that exhibit differing outputs. Typically there are very few (\eg,
1--3) cases that require human inspection to confirm and fix, a
process that typically requires only a few minutes per contract. Since
contracts are not patchable after deployment, we believe that
\toolname is useful auditing tool: it flags about $7$--$11\%$ of
contracts it analyzes, reporting less $1$--$3$ witnesses for analysts
to inspect, over half of which have yielded subtle EO bugs.

%
%
% The third observation from contracts-as-concurrent-objects analogy
% suggests a way to design an efficient procedure for detecting EO bugs,
% by drastically reducing the search space of event orderings.
% %
% %
% Our tool pre-processes contract implementations to discover their HB
% structure by examining which function pairs commute to avoid
% combinatorial search. Combined with a symbolic analysis of the event
% values, this approach reduces the analysis time for a contract from
% weeks down to a few minutes.

% %
% To achieve our goals, we propose a new problem formulation and a
% principled dynamic analysis strategy for finding EO bugs.
% %
%
% Ethereum smart contracts can have multiple entry points, and functions
% in a contract can be executed in many different orders under the
% control of the adversarial users or mining network. The core
% hypothesis of this work is that developers do not anticipate the
% behaviour resulting from such re-ordering, leading to subtle
% vulnerabilities which are challenging for human developers to
% identify. Smart contracts carry virtual coins (Ether) and cannot be
% patched after deployment; hence, their security analysis
% pre-deployment is of particular importance.

%
%
%
% Further, we observe that contracts are written expecting certain pair
% of events to execute atomically, formalized as a ``linearizability''
% property of event orderings. Only non-linearizable orderings yielding
% outputs different to all linearizable ones result in EO bugs.

\section{Motivation}
\label{sec:motivation}

\subsection{Ethereum Smart Contracts}

Smart contracts in Ethereum are identified by addresses. An Ethereum
user invokes a particular function of a smart contract by creating a
signed transaction to the contract's address.  
%The contract can take explicit arguments. 
The transaction specifies which function is being
executed and its arguments.  The user submits the transaction to the
Ethereum network, and at some point in the future, a miner in the
network chooses to process the transaction. Accordingly, the miner
takes the current state of the contract from the blockchain, executes
the called function, and stores the updated state of the contract back
into the blockchain.

It is entirely possible for two users to submit transactions that
interact with the same contract at the same time. Neither of the users
knows which transactions will first be processed by the miners.  What
is guaranteed is that if a user's transaction is incorporated into the
blockchain, then its effect will be reflected atomically.  In other
words, a miner will \emph{not} execute part of the first transaction,
switch to running the second at some intermediate contract state, and
then return to finish off the remaining computation in the first
transaction.  It is easy to assume that this \emph{transaction
  atomicity} removes the need to reason about the concurrent execution
environment, but this is not the case, as we explain next.

\subsection{Event-Ordering Bugs}
\label{sec:eobugs}
%\vspace*{0.2cm}

Contracts can be seen as objects with mutable state and set of interfaces
for users to access it. As explained, their interfaces can be invoked
by many users simultaneously; the order in which the ``calls'' will be
invoked is determined entirely by the mining network. Phrased this
way, one can readily see the analogy between concurrent data
structures in traditional programming languages and smart contracts.
Traditional programming languages provide programming abstractions to
allow concurrent data structures to ensure certain desired
properties. We explain two properties here
which are desirable by smart contracts, violations of which result in
errors defined as event-ordering (or EO) bugs.

\paragraph{\Lin}
\Lin is a well-known property used in concurrent data
structures~\cite{Herlihy-Wing:TOPLAS90}. Loosely stated, it means that
an operation should appear to execute atomically or instantaneously
from each user's view. A simple example of this is the operation of
incrementing a global counter, shared between two threads in a
classical programming language. The counter implementation reads the
counter value, increment by one, and write the result back. If two
users invoke simultaneously when the counter value is zero, a poor
implementation may result in a final value of $1$ rather than $2$, if
reads of two user requests execute before both the writes; the last
write would overwrite the other. Linearizability guarantees that no
interleavings would result in such outcome.  Specifically, the result
should match the sequential execution of one request being served
after another, either order being legal.  Traditional language and
hardware instructions, such as atomic swap and compare-and-exchange
operations, allow programmers to achieve linearizability through
mutual exclusion.

%\begin{multicols}{2}
%\blindtext
\begin{figure}[t]
\begin{center}
\centering
\begin{tabular}{cc}
%\begin{minipage}{0.50\linewidth}
\begin{lstlisting}[mathescape=true,language=Solidity,basicstyle=\footnotesize\ttfamily]
contract Casino {
  ...
  function bet() payable {
    // make sure we can pay out the player
    if (address(this).balance < msg.value * 100 ) throw; $\label{code1:betcheck}$
    bytes32 oid = oraclize_query(...); // random $\label{code1:calloracle}$
    bets[oid] = msg.value; $\label{code1:storebet1}$
    players[oid] = msg.sender; $\label{code1:storebet2}$
  } $\label{code1:endbet}$

  function __callback(bytes32 myid, string result) $\label{code1:callback}$
    onlyOraclize onlyIfNotProcessed(myid) {
    ...
    if (parseInt(result) % 200 == 42) $\label{code1:wincheck}$
    players[myid].send( bets[myid] * 100 ); $\label{code1:payout}$
  }
  ...
}
\end{lstlisting}
%\end{minipage}
\end{tabular}
\end{center}
\caption{Contract \code{Casino} with a \lin violation}
%\caption{\texttt{Buggy} contract.}
\label{fig:casino}
\end{figure}
%\end{multicols}

Ethereum contracts desire linearizability, but the transaction
atomicity provided by the platform does not guarantee it.
Consider the \code{Casino} snippet in
Figure~\ref{fig:casino}, which was simplified from a real
game-of-chance smart contract.
\code{Casino} accepts bets from one or more players and, with
200-to-1 odds (Line~\ref{code1:wincheck}), repays winners 100 fold
(Line~\ref{code1:payout}). The \texttt{Casino} aims to be honest, so it rejects
any bet that it would be unable to honor if won
(Line~\ref{code1:betcheck}).  The fairness of any game of chance
depends crucially on how random values are generated.  \code{Casino}
utilizes a trusted off-chain random number generator which is invoked
by the \Oraclize API~\cite{oraclize} query
(Line~\ref{code1:calloracle}).  The random-number oracle is not
actually queried in Line~\ref{code1:calloracle}; instead calling
\code{oraclize_query} does two things: generates the unique
transaction id tag \code{oid} and notifies a trusted off-chain
monitor (the \Oraclize service) that \code{Casino} wishes to query
the random-number oracle for transaction \code{oid}.  Due to the
semantics of Ethereum, the off-chain monitor will be notified only
once \code{Casino}'s code has finished running
(Line~\ref{code1:endbet}).  Before this occurs, \code{Casino} stores
transaction-related information for its later retrieval
(Lines~\ref{code1:storebet1}--\ref{code1:storebet2}).  Later, once the
off-chain oracle has been queried, the off-chain \Oraclize service will
make a fresh call back into the Ethereum chain at \code{Casino}'s
callback function \code{__callback} (Line~\ref{code1:callback}).

The \code{__callback} function ``returns'' asynchronously.
After an initial bettor initiates an oracle query, other bettors can
place their bets while the off-chain oracle is queried.  These further
wagers will initiate further oracle queries, and depending on the
behavior of the off-chain oracles, their corresponding callbacks may
not be invoked in the same order as they are called. The designers of
the \Oraclize API are aware of this, which is why each
transaction is given a unique ID that is both returned from
\code{oraclize_query} (Line~\ref{code1:calloracle}) and passed to
the callback (the \code{myid} parameter in
Line~\ref{code1:callback}),  thereby ``pairing'' the two.
Failing to pair callbacks can lead to previously-published
vulnerabilities~\cite{Sergey-Hobor:WTSC17}, but this error is avoided
by \code{Casino} by the use of \code{myId} in
Line~\ref{code1:payout}, which carefully disambiguates bets from
multiple users.

Even though the call/return are correctly paired, there is a bug that
can occur when multiple players place bets concurrently.  Suppose that
the contract currently has 100 Ether and that two players wish to bet
1 Ether.  Consider the following execution of functions: {\small{$\langle
\texttt{bet}_1$; $\texttt{bet}_2$; $\texttt{\_\_callback}_1$;
$\texttt{\_\_callback}_2 \rangle$}}, where the subscript denotes the
paired identifier (\code{oid}/\code{myId}) in the
\Oraclize interface. Both bets are accepted
(Line~\ref{code1:betcheck}), but if both bets win
(Line~\ref{code1:wincheck}), player two will not be paid in full.  A
fairer \code{Casino} implementation should have considered all
pending bets when determining if it can accept another.

This example highlights exactly why {\em linearizability} is
desirable---the call-return of \Oraclize calls for two users
should appear to execute sequentially (or atomically). This contract
yields differing outputs if the responses of callbacks are received
out-of-order. The ordering presented above yields an insufficient
balance after paying off player 1 (Line~\ref{code1:payout}).  In the
alternative ordering {\small{$\langle \texttt{bet}_1$;
$\texttt{\_\_callback}_1$; $\texttt{bet}_2$; $\texttt{\_\_callback}_2
\rangle$, $\texttt{bet}_2$}} will decline the second bet due to the
check on Line~\ref{code1:betcheck}, making \code{Casino} fair.

%\begin{multicols}{2}
%\blindtext
{\setlength{\belowcaptionskip}{-15pt}{
\begin{figure}[t]
\begin{center}
\centering
\begin{tabular}{cc}
%\begin{minipage}{0.50\linewidth}
\begin{lstlisting}[mathescape=true,language=Solidity,basicstyle=\footnotesize\ttfamily]
contract IOU {
 // Approves the transfer of tokens
 function approve(address _spender, uint256 _val) { $\label{code2:approve}$
   allowed[msg.sender][_spender] = _val;
   return true;
 }
 // Transfers tokens
 function transferFrom(address _from, address _to,$\label{code2:transferFrom}$
                       uint256 _val) {
   require(
     allowed[_from][msg.sender] >= _val
     && balances[_from] >= _val
     && _val > 0);
   balances[_from] -= _val;
   balances[_to] += _val;
   allowed [_from][msg.sender] -= _val; $\label{code2:updatebalance}$
   return true;
 }
}
\end{lstlisting}
%\end{minipage}
\end{tabular}
\caption{Contract \code{IOU} with a \syn violation}
\label{fig:ERC20}
\end{center}
\end{figure}
}}
%\end{multicols}

\paragraph{\Syn}
\Syn is the idea that two processes order their execution, such that
certain operations execute in a desired order. The classical example
of this is the producer-consumer problem for a shared queue, where a
producer process should not write to a full buffer, and a consumer
process does not read out of an empty one. Typically, \syn is
implemented with abstractions such as semaphores in traditional
languages.

Smart contracts implicitly desire to order multiple user requests in a
particular sequence, but sometime fail to. As an example,
Figure~\ref{fig:ERC20} shows a shortened version of a contract that
implements \erc-compliant tokens in Ethereum dubbed ``IOU''s.  As of
today, \erc-compliant contracts manage tokens valued in the billions
of USD.  The snippet in Figure~\ref{fig:ERC20} allows an owner ``$O$''
of IOUs to delegate control of a certain \code{_val} of IOU tokens
to a specified {\small{$\texttt{\_spender}$ ``$S$'}}' (\eg, an expense
account).  $O$ calls the \code{approve} function
(Line~\ref{code2:approve}) to allocate \code{_val} IOU tokens to
$S$, and the function {transferFrom} allows $S$ to send a portion
(\code{_val}) of the IOU allocation to an address of $S$'s choice
(\code{_to}).
%Notice that on Line~\ref{code2:updatebalance}, the balance of  $S$ is decremented by the now-sent IOUs.
%to account for the tokens now sent.
%
The approver is allowed to update the allocation any time: for
instance, $O$ may initially approve $300$ IOU and later reduce the
amount to $100$ by calling \code{approve} again. Although this may
seem like a reasonable idea, as pointed out on public forums~\cite{erc20bug}, this
contract has undesirable semantics, since $S$ can execute a
\code{transferFrom} between the two calls to \code{approve},
thereby spending the first allocation of $100$ and then having another
$100$ to spend. In other words, two different executions of calls to \code{approve} and \code{transferFrom} lead to different outcomes:

\begin{center}
\vspace{8pt}
{\small{
\begin{tabular}{l@{\ \ }l}
\multicolumn{1}{c}{Execution 1} & \multicolumn{1}{c}{Execution 2}
\\
\hline
{\lstinline[language=Solidity,basicstyle=\footnotesize\ttfamily,mathescape]!approve$_O$(S, 300)!}
&
{\lstinline[language=Solidity,basicstyle=\footnotesize\ttfamily,mathescape]!approve$_O$(S, 300)!}
\\
\lstinline[language=Solidity,basicstyle=\footnotesize\ttfamily,mathescape]!approve$_O$(S, 100)!
&
\lstinline[language=Solidity,basicstyle=\footnotesize\ttfamily,mathescape]!transferFrom$_S$(O, S, 100)!
\\
\lstinline[language=Solidity,basicstyle=\footnotesize\ttfamily,mathescape]!transferFrom$_S$(O, S, 100)!
&
\lstinline[language=Solidity,basicstyle=\footnotesize\ttfamily,mathescape]!approve$_O$(S, 100)!
\\
%\lstinline[language=Solidity,basicstyle=\small\ttfamily,mathescape]!transferFrom$_B$(A, C, 100)!
%&
%\lstinline[language=Solidity,basicstyle=\small\ttfamily,mathescape]!transferFrom$_B$(A, C, 100)!
%\\
\hline
$S$ \emph{has spent 100 and} & $S$ \emph{has spent 100 and}
\\
\emph{can spend 0 more} & \emph{can spend 100 more}
%
%\\\hline
\end{tabular}
}}
\vspace{5pt}
\end{center}

The community has proposed a fix that corresponds directly to the
desired synchronization property: forbidding a change an allocation
once made~\cite{erc20bug}.  The recommendation ensures that all calls
to \code{transferFrom} by all users should be permitted strictly after
all \code{approve} have happened. Newer \erc contracts have deployed
this fix and behave more reasonably if such order is enforced, since
the second \code{approve} is rejected in the second ordering.

% \begin{enumerate}

% \item
% %
% \lstinline[language=Solidity,basicstyle=\small\ttfamily,mathescape]{approve$_A$(B, 200)},
% \lstinline[language=Solidity,basicstyle=\small\ttfamily,mathescape]{approve$_A$(B, 100)},
% \lstinline[language=Solidity,basicstyle=\small\ttfamily,mathescape]{transferFrom$_B$(A, C, 200)},
% \lstinline[language=Solidity,basicstyle=\small\ttfamily,mathescape]{transferFrom$_B$(A, C, 100)},

% \item
% %
% \lstinline[language=Solidity,basicstyle=\small\ttfamily,mathescape]{approve$_A$(B, 200)},
% \lstinline[language=Solidity,basicstyle=\small\ttfamily,mathescape]{transferFrom$_B$(A, C, 200)},
% \lstinline[language=Solidity,basicstyle=\small\ttfamily,mathescape]{approve$_A$(B, 100)},
% \lstinline[language=Solidity,basicstyle=\small\ttfamily,mathescape]{transferFrom$_B$(A, C, 100)}.

% \end{enumerate}

\section{Overview}

We define a class of bugs which we call \emph{event-ordering bugs},
generalizing the examples presented in Section~\ref{sec:motivation}.
We explain the inherent path and state space exploration complexity,
and propose our design to address these systematically.

\subsection{Problem}

A contract function can be invoked by an external transaction, an
internal call from another contract, or via an off-chain asynchronous
callbacks. We call these invocations as {\em events}. Under a received
event, a contract executes atomically, which we call as {\em run}. A
sequence of events invokes a sequence of contract runs, which is
referred to as a {\em trace} in this paper. Each run of a contract can
modify its Ether balance and global state which is stored on the
blockchain, as well as generate new transactions that transfer Ether
or call other contracts. We say that the {\em output} of a run or a
trace are values of the contract balance, state, and any resulting
transactions. We define these terms more precisely in
Section~\ref{sec:formal}.

An event-ordering (EO) bug exists in a contract if two different
traces, consisting of the same set of concrete events, produce
different outputs. These bugs may arise as a result of violations of
both \lin and \syn properties. We seek to check if a given smart
contract has an event-ordering (EO) bug.

Note that our problem formulation is more general than previous works
which introduced a sub-class of synchronization mistakes called
transaction ordering bugs (TOD)~\cite{Luu-al:CCS16} (see
Section~\ref{sec:oy} for details).  TOD bugs capture only a pair of
events and are restricted only to differences in contract's balance. Our EO bug
formulation is also orthogonal to the concept of re-entrancy bugs,
such as the infamous DAO, since it does not correspond to re-ordering
of events. A re-entrancy bug occurs when a contract exhibits different
outputs under two different values of a {\em single} event, one of
which outputs a transaction which could subsequently cause a critical
function to re-enter itself. Re-entrancy bugs are detected by several
tools that analyze runs from different values for a single
event~\cite{Luu-al:CCS16,Grossman-al:POPL18}.

{{
\begin{figure}[t]
\centering
\includegraphics[width=\linewidth]{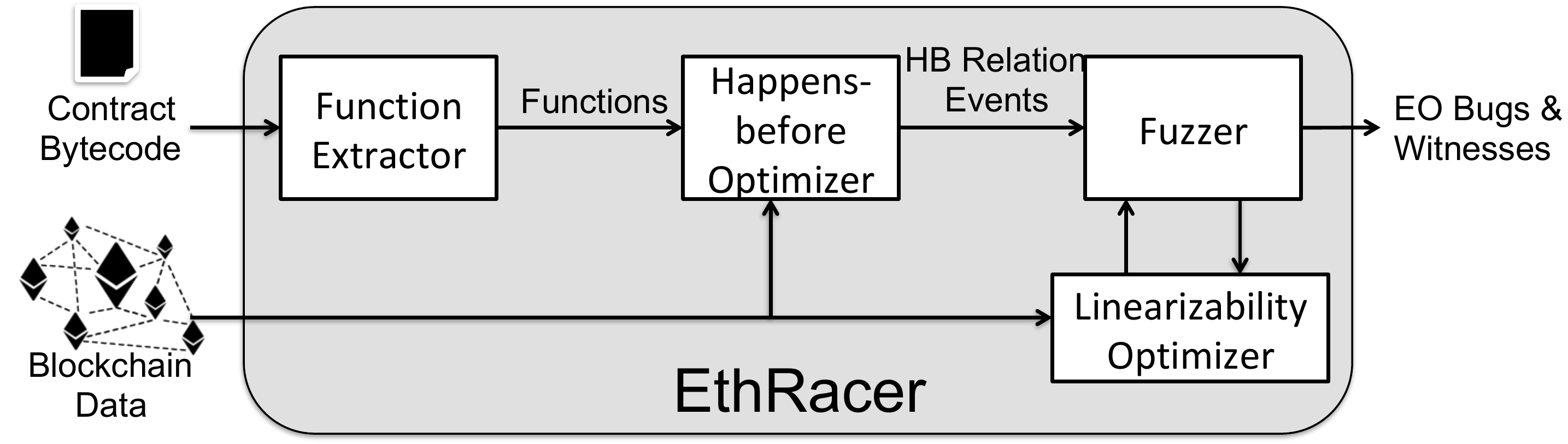}
\caption{Main components of \toolname. \todo{Are you sure there is no
    interaction between the main pipeline and \oraclize? Doesn't the
    former take events from the latter?}}
\label{fig:tool}
\end{figure}
}}

Our goal is to efficiently check if re-ordering of input events of a
contract across multiple transactions leads to differing outputs. We
design a novel analysis technique and an automatic tool called
\toolname to find such contracts. As shown in Figure~\ref{fig:tool},
\toolname works directly on the EVM bytecode of Ethereum smart
contracts, given as input along with the public blockchain data. 
When an EO violation is detected, the output of \tool is minimally two
traces, consisting of at least two distinct events, which concretely
exhibit different outputs when executed on the Ethereum virtual
machine (EVM) from the given blockchain state.

\paragraph{Intent vs. Bugs}
Unlike vulnerabilities arising due to language design, such as memory
safety, the bugs uncovered in the present work tend to arise from
logical errors between a contract's intended behavior and its actual
implementation.  Smart contracts are usually deployed anonymously and
without public documentation; the true motivations for their
deployment can be very obscure.  It is possible that our flagged test
cases detected as having output differences are desired by the
developer.  Without documentation, to say nothing of a formal
specification, it is difficult to give a precise definition for
``intended behavior'', forcing us to fall back on the notion of ``know
a bug when we see it''~\cite{stewart:64}.  In reality, the debate
between intention and formal behavior is sometimes so contentious that
it is resolved in the community by agreeing on a so-called hard
fork~\cite{theDao}.

Therefore, our focus in this work is merely on efficient analysis
techniques that minimize the number of test configurations the
developer or potential user has to manually inspect, before the
contract is deployed irrevocably.

\subsection{Our Solution}

We take a dynamic testing approach to finding EO bugs. This is
primarily motivated by our goal to produce concrete witnesses that
human analysts can replay to inspect and confirm EO bugs with no false
violations. The technical barrier to finding reordering bugs is that
the path and the state space of the analysis blows up combinatorially.
To illustrate this, we consider a baseline solution that (say) can
reason about the behavior of the contract under different values of a
single input event. One effective state-of-the-art technique for such
analysis is based on dynamic symbolic execution (DSE), implemented by
many existing analyses for
contracts~\cite{Nikolic-al:Maian,Kalra-al:NDSS18,Krupp-Rossow:USENIX18}.

In concept, such a DSE engine can enumerate different paths and input
values (symbolically) in the program, starting from one event entry
point function, and check if two different paths could lead to
different outputs. However, this baseline solution does {\em not}
address the class of EO bugs sought directly. EO bugs are a result of
two or more events with possibly different entry points, leading to
two different paths and outputs. Therefore, even with a powerful DSE
engine, checking the space of $N$ events requires reasoning about
traces with all $N!$ permutations of these events. Different traces
can result in different intermediate values global variables or
states, leading to different paths. Enumerating such a path space is
prohibitively expensive; for instance, even a (relatively short)
sequence of calls to the contract with (say) 20 functions callable via
events can have millions of traces to inspect.\footnote{\Eg, for a typical
  contract with 20 functions, each with 2 different inputs, the number of traces composed of 6 events is
  $C_{(20\cdot 2\cdot 6)}^{6} \approx 2^{38}$.} Therefore, we seek
techniques that can eliminate a large part of this path and state space
exploration, making search tractable.

\paragraph{Key Ideas}
The first observation is that contract functions are often written in
a way that have a certain intended order of {\em valid} execution.
Executing them in a different order simply results in the contract
throwing an exception. This can be captured as a partial-order between
events or as a happens-before (or $\Hb$) relation. Two events $e_1$
and $e_2$ are in a $\Hb$ relation if executing them in one order produces
a valid trace while the other is invalid. The insight is that if we
find pairs of events $e_1$ and $e_2$ that are in $\hb{e_1}{e_2}$
relation, {\em all}\footnote{The savings can be exponential, since
  there are exponentially many permutations with a shared pair of
  re-ordered functions.}
permutations of events involving these in an invalid order will result
in exception---therefore, these are redundant to test repeatedly. To
maximize the benefit of this observation, we augment the baseline DSE
to recognize events which are in the $\Hb$ relation.  The symbolic
analysis reasons about a large set of event values encoded as symbolic
path constraints, which helps to identify partially-ordered events
much better than fuzzing with concrete values. We present our novel
and efficient algorithm to extract a new notion of $\Hb$ relations 
we define in Section~\ref{subsec:extracthb}.

Our second observation is that certain events produce the same outputs
irrespective of which order they appear in any trace. A simple, but
highly effective, optimization in our experiments is to recognize
events runs that do not write or read global state. Such %{\em pure} (with no side-effects) 
events can be arbitrarily re-ordered without changing
outputs, so testing permutations that simply re-order these is
unnecessary. Similarly, events that do not write to any shared global
state can be arbitrarily re-ordered. Hence, they are also not considered 
for extracting HB relation.

% A closely related observation is that events that have
% no shared variables written to (have disjoint write-sets). cannot influence each other
% and thus are trivially not in $\Hb$; therefore, their symbolic analysis
% for extracting $\Hb$ can be eliminated.

The final key observation deals primarily with optimizing for \lin
violations. 
% Unlike concurrent objects, contracts are not accompanied
% by a specification of correctness properties; yet \tool hopes to
% extract EO bugs useful to practitioners.  
This optimization aims to prioritize the search for event traces involving
asynchronous callbacks and reduces flagging EO violations that may be
benign. 
%Recall that our example of contracts often implicitly desire certain operations to execute atomically. 
If we only consider traces where each \Oraclize and its matching
callbacks are sequentially or atomically executed, these orders likely
yield intended (benign) outputs. Such traces, in which events
corresponding to different user requests are not ``out-of-order'', are
defined as {\em linearizable}. 
%Linearizable traces encode a notion of benign ordering, leading to intended outputs; 
We explain this concept of linearizability more precisely as applied to
contracts in Section~\ref{subsec:call-return}. The task of analysis reduces to
finding if there exist traces that do {\em not} produce outputs same as that
of some linearizable trace. Orders which result in the same output as a
linearizable traces are likely not bugs (can be suppressed in reporting), and
those which produce differing outputs are flagged as an EO bug. So, our search
procedure prioritizes constructing linearizable traces first and then
symbolically analyzing for event runs which may yield different outputs to
those linearizable traces.

In summary, \toolname uses the baseline DSE technique to prune and
prioritize its search in the path space of runs in multi-event traces.
The final concrete values generated from DSE are run on the Ethereum
Virtual Machine (EVM) from the provided blockchain state. As a result,
in practice, testing for EO bugs takes about $18.5$ minutes for $95\%$
with \toolname with traces of length $6$, and can be configured to
test for longer traces arbitrarily. We detail the tool design details
in Section~\ref{sec:motivation}.

\section{\toolname Design}
\label{sec:tool}

The notion of concurrency in contracts requires a careful mapping to
the familiar concepts of concurrency. We begin by recalling the
contract concurrency model and by precisely defining terminology used
informally thus far. We utilize the notation used in a distilled form
of Ethereum Virtual Machine (EVM) semantics called \elite
calculus~\cite{Luu-al:CCS16}.  Then, we move to design detail of our
tool \codename.

\subsection{Contract Concurrency Model \& Terminology}
\label{sec:formal}

% An Ethereum contract can be invoked via a transaction specifying a
% function in it. Contracts can have multiple entry points, or functions
% which can be externally invoked. Each call to an entry point, via an
% instruction, is called an {\em event}. An event, intuitively, consists
% of the function name being invoked, and values of the inputs it reads.
% 
% 
%  Two
% transactions creating events $e_1$ and $e_2$ respectively, can be
% ordered in two different ways by two miners. The order $[e_1, e_2]$ is
% processed such that the output of event $e_1$ is used in $e_2$, and
% the output of $e_2$ will proposed as final, since events execute
% sequentially on a miner's computer. However, another miner may order events
% $[e_2, e_1]$ leading to a different output. It is this non-determinism
% that can be viewed as two possible ``schedules'' of the events $e_1$
% and $e_2$. Eventually, the miners resolve the non-determinism to a
% single agreed schedule and resulting output value; but this is only
% determined by consensus protocol which is not deterministic. In this
% paper, if different ordering of events can lead to different outputs,
% we say that the contract has an EO bug.
%\subsection{Formalization}
%

Recall that each event executes atomically and deterministically in
Ethereum~\cite{Gavin-al:yellow-paper}. A miner runs the contract
function, conceptually as a single thread, with provided inputs of an
event until it either terminates or throws an
exception.\footnote{Here, we do not draw a difference between
  different origins of exceptions, \ie, those raised programmatically
  or those triggered by insufficient {\em gas}.}  If it terminates
successfully, all the changes to the global variables %and other outputs (including transactions) 
of that execution are committed to the blockchain; if an
exception is thrown, none of the changes during the execution under
that event are committed %as output 
 to the blockchain. In this sense,
the execution of one event is atomic.  The source of concurrency lies
in the non-deterministic choices that each miner makes in ordering
transaction in a proposed block.

\paragraph{Contract states and instances}
We recall the definitions of global blockchain state, contracts and messages 
from \elite calculus by Luu~\etal~\cite{Luu-al:CCS16}.

A global blockchain state $\bstate$ is encoded as a finite partial
mapping from an account $\Id$ to its balance and contract code $M$ and
its \emph{mutable} state, mapping the field names $\fld$ to the
corresponding values, which both are optional (marked with~``?'') and
are only present for contract-storing blockchain records.
We refer to the union of a contract's field entries $\many{\fld
  \mapsto v}$ and its balance entry $\Bal \mapsto z$ as a
\emph{contract state}~$\cstate$,\footnote{We will also overload the
  notation, referring to a state $\cstate$ of a contract $\Id$ in a
  blockchain state $\bstate$ as $\cstate = \bstate[\Id]$, thus,
  ignoring its $\Code$ component.}
and denote a triple $c = \angled{\Id, M, \cstate}$ of a contract with
an account $\Id$, the code of which is $M$ and state is $\cstate$, as
\emph{contract instance}~$c$.

{\small{
\[
\begin{array}{l@{\ \ }c@{\ \ }l}
\bstate & \edeff & \many{\Id \mapsto \set{\Bal : \Nat;~\Code? \mapsto
                     M;~ \many{\fld? \mapsto v}}}
\end{array}
\]
}}

\noindent
Messages (ranged over by $m$) are encoded as mappings from identifiers
to heterogeneous values. In particular, each message stores the
identity of its $\Sender$ and destination ($\To$), the amount $\Value$
of Ether being transferred (represented as a natural number), a
contract function name to be invoked, as well as auxiliary fields
($\Data$) containing additional arguments for a contract
function, which we will be omitting for brevity.

{\footnotesize{
\[
\!\!\!
\begin{array}{l@{\ \ }c@{\ \ }l}
 m & \edeff & \set{\Sender \mapsto \Id; \To \mapsto \Id';~\Value :
                                \Nat;~ \fname : \String~; \Data \mapsto ...}
\end{array}
\]
}}

\vspace{-10pt}
\noindent
We will refer to a value of an entry $x$ of a message~$m$~as~$m.x$.

\paragraph{Contract Events} 
The notion of events captures external inputs that can force control
into an entry point of a contract, and defined  as follows.

\vspace{3pt}
\begin{definition}[Events]
\label{def:event}
An \emph{event} $e$ for a contract instance $c = \angled{\Id, M,
  \cstate}$ is a pair $\angled{\fn, m}$, where $\fn$ is a name of a
function, defined in $M$, and $m$ is a message passed containing
arguments to $\fn$, such that $\fn = m.\fname$.\footnote{In EVM,
  function names are not different from other fields of the message,
  but here we make this separation explicit for the sake of clarity.}
\end{definition}
\vspace{3pt}
Below, we will often refer to the $m$-component of an event as its
\emph{context}.  Our contract events are \emph{coarse-grained}: they
are only identified by entry points and inputs to a specific contract.
Event executions of a contract may invoke other contracts, which only
return values as parameters. Externally called contracts may modify
their own local state, but such external state is not modelled.

\paragraph{Event runs and traces}
The \emph{run} of an event $e$ at a contract instance $c =
\angled{\Id, M, \cstate}$ brings it to a new state $\cstate'$, denoted
$\cstate \estep{e} \cstate'$. A sequence of events is called an {\em
  event trace}. An evaluation of contract at instance $\cstate_0$ over
an event trace $h$ yielding instance $\cstate_n$ is denoted as
$\cstate_0 \tstep{h} \cstate_n$ $= \cstate_0 \estep{e_1} \ldots
\estep{e_n} \cstate_n$, which can be obtained by executing all events
from $h$ consecutively, updating the global blockchain state
correspondingly between the steps. Some evaluations may halt
abnormally due to a runtime exception, denoted as $\cstate \tstep{h}
\exn$. We call such $h$ {\em invalid traces} (at~$\cstate$).
Conversely, valid traces evaluate at instance $\cstate_0$ without an
exception to a well-formed state $\cstate_n$, which is implicit when
we write $\cstate \tstep{h} \cstate_n$. We now define an event
ordering bug.

% 
% \begin{definition}[Valid event history]
% \label{def:hist}
% For a blockchain state $\bstate$, contract instance $c = \angled{\Id,
%   M, \cstate_0}$, such that $\bstate[\Id] = \cstate_0$, and an event
% $e = \angled{\fn, m}$, such that $m.\To = \Id$, we define the event
% execution relation $\estep{e}$ as follows:
% %
% an event history $h = [e_1, \ldots, e_n]$, represented as a finite
% list of events (such that $\forall i, e_i.m.\To = \Id$), is
% \emph{valid}, iff 
% \end{definition}
% % In a blockchain setting, where the scheduling of such events is not
% %predetermined upfront, there is a possibility for any ordering of
% %such invocations.

\begin{definition}[Event-ordering bug] 
\label{def:eobug}
For a fixed contract instance $c = \angled{\Id, M, \cstate_0}$, a
blockchain state $\bstate$, such that $\bstate[\Id] = \cstate_0$, a
pair of valid event traces $h = [e_1, \ldots, e_n]$ and $h' =
[e'_1, \ldots, e'_n]$ constitutes an event-ordering bug \Iff
\begin{itemize}
\item $h'$ is a permutation of events in $h$,
%\item $\forall e_i, e_j \in \set{e_1, \ldots, e_n}, \neg \hb{e_i}{e_j}$, and
\item if $\cstate_0 \tstep{h} \cstate_n$ and $\cstate_0 \tstep{h'}
  \cstate'_n$, then $\cstate_n \neq \cstate'_n$.
\end{itemize}
\end{definition}

Due to the coarse-grained nature of events, targeting
multi-transactional executions, our concurrency model does not capture
reentrancy bugs~\cite{reentrancy}. 
Furthermore, for the sake of tractability, the definition is only
concerned with a single contract at a time, even though involved
events may modify state of other contracts.
This means that our notion of EO bugs only captures the effects on a
\emph{local} state of a contract in focus and won't distinguish
between states of other contracts.\footnote{One could generalize
  Definition~\ref{def:eobug}, allowing to catch bugs resulting in
  transferring money to one beneficiary instead of another, without
  any local accounting. In this work, we do not address this class of
  \emph{non-local} EO bugs.}

\noindent

\subsection{The Core Algorithm}

\toolname basic algorithm systematically enumerates all traces of up to
a bounded length of $k$ events (configurable), for a given smart
contract. One could employ a purely random dynamic fuzzing approach to
generating and testing event traces.  However, even for a single entry
point, different input values may exercise different code paths. In
contrast, symbolic execution is a useful technique to reason about
each code path, rather than enumerating input values, efficiently. One
could consider a purely symbolic approach that checks properties of
code paths encoded fully symbolically, to check for output
differences. \tool uses an approach that combines symbolic analysis of
contract code and randomized fuzzing of event trace combinations in a
specific way to find EO bugs.

% The tradeoff between a purely symbolic and purely concrete fuzzing
% approach is the empirical performance of the two. \tool spends $15$
% out of $18.5$ minutes on average (see Section~\ref{sec:performance})
% on symbolic analysis, and the remaining on concrete fuzzing. Our
% design balances the benefits of an expensive symbolic analysis with
% that of undirected, but much faster, concrete fuzzing.

% uses a set of principled ideas to reduce its search space
% further.  Instead of enumerating event contexts (parameters of invoked
% functions), \tool uses standard symbolic execution techniques to
% generate symbolic contexts, and use one concrete instance for each
% symbolic context. After these techniques, \tool reduces the search
% space significantly and performs fuzzing to find differing outputs.

\paragraph{Symbolic Event Analysis}
\tool first performs a syntactic analysis to recover the public functions in the
bytecode of the contract. These functions are externally
callable, and thus, these are entry points for each event. Next, \tool
employs standard dynamic symbolic execution technique to reason about
the outputs of each event separately. For each event $e_i$, the
analysis marks the event inputs as symbolic and gathers constraints
down all paths starting from the entry point of $e_i$.  Modulo
implementation caveats (as detailed in Section~\ref{sec:symbolic-execution}),
our analysis creates constraints that over-approximate the set of
values that drive down a specified path. Note that since the symbolic
analysis aims to over-approximate the feasible paths concretely
executable under an event, it does not need to always check the
feasibility of each path exactly using SMT solvers in this phase.
As a result of this analysis, for each event $e_i$, we obtain a vector
of symbolic constraints (as SMT formulae) $\vec{S_i}$ that encode path
constraints for the set of execution paths starting from the entry
point of $e_i$. The pairs $\angled{e_i, \vec{S_i}}$ is referred to as
a {\em symbolic event}.

% static symbolic analysis of the entire code executable
% under an event. In this static symbolic analysis phase it extract the
% symbolic variables potentially read by the code (before clobbering)
% from the input transaction data along all paths of the
% contracts. These symbolic variables constitute the inputs or
% invocation {\em context} of the event.  We explain how this
% done later.  This phase provides us the vector of symbolic context
% variables $\vec{V}$ for each event starting at a function $f$.  We also
% keep the path constraints gathered on each input variable in $\vec{V}$
% in a {\em constraint map} for later analysis.
% A contract can have multiple entry points, and hence multiple symbolic
% events. \tool aims to create all ordering (or histories) of $k$
% symbolic events and test the behavior of the contract under all such
% orderings. To do so, \tool uses the constraint map to extract the
% symbolic constraints on each context variable in $\vec{V}$ of each
% symbolic event. 

\paragraph{Concretization}
Conceptually, symbolic events are concretized in two separate steps.
The first concretization step is standard in DSE system, where
concretization aids path exploration. Specifically, the feasibility of
each path constraints in $\vec{S_i}$ using an off-the-shelf SMT
solver. We eliminate symbolic path constraints for paths that become
infeasible.  The SMT solver finds satisfying concrete input values for
feasible paths.  By executing these inputs on the blockchain state
given as input to \tool, it can find concrete values of global
symbolic variables read before being clobbered, call targets, and
transaction input contexts.  These values can concretize symbolic
constraint in $\vec{S_i}$.  This concretization makes the symbolic
constraints less general, but allow pruning away a lot of false
positives that due to assuming infeasible paths or values of symbolic
inputs. All our prioritization and pruning optimization outlined later
in this Section use the outputs of this concretized symbolic execution
for analysis.

The second concretization step aims to create one or more concrete
value for each path that remains in $\vec{S_i}$.  The goal of this
step is enumerate all the traces $k$ concrete events long, and
concretely fuzz the contract with these concrete event traces.  We use
an SMT solver to find value assignments for all symbolic variables in
each element of $\vec{S_i}$. These concrete values give us a set of
{\em concrete events}. Concrete events can be directly executed on the
Ethereum Virtual Machine (EVM).

\paragraph{Fuzzing with Concrete Events}
The fuzzing starts with the random trace of $k$ concrete events
distinct entry points, and its creates combinations that permute all
pairs of events, then all triples, quadruples, and so on. If two
traces tested result in differing outputs, we flag the contract as
having an EO bug. A pair of concretely tested traces that exhibit
diverging outputs and are called a {\em witness pair}, and reported by
\toolname for evidence of EO violations. To reduce the effort of the
human analyst, \tool performs a minimization step on witness pairs. If
the removal of a concrete event from both traces in a witness pair
does not impact the outputs, such an event is redundant and can be
removed. Proceeding this way, a {\em minimized witness pair} is
finally reported.

\paragraph{Prioritizing and Pruning}
We devise three new optimization strategies, which cut down the
fuzzing space exponentially:

(a) \tool uses its symbolic analysis to infer a $\Hb$ relationship
between symbolic events after the first concretization step. If an
concrete event $e_1$ is before $e_2$ in $\Hb$, it implies that
re-ordering them will lead to an invalid trace throwing an exception.
Our tool analyzes a set of such concrete events symbolically,
which is much faster than learning such relationships through concrete
fuzzing. This procedure is described in Section~\ref{sec:fuzzing}.
Note that this optimization can have an exponential reduction in
concrete fuzzing phase. Learning an $\Hb$ relation between even two
events can eliminate generating and testing half of the $N!$
combinations.\footnote{By symmetry, half the combination have (say)
  event A before B.} The second concretization step, therefore,
generate concrete events that respect the $\Hb$ relations.

(b) During its symbolic analysis, \tool  identifies %pure functions,
%\ie 
the functions that do not read or write any global variables. \tool further does not consider events from such functions 
%prunes away traces that only reorder certain pure functions, \ie events that
because they cannot interfere with any other event. While this is a simple 
observation, in practice, it helps optimization even further by
cutting down what requires symbolic analysis.

(c) When handling asynchronous callbacks (which are represented by
events in our model), the fuzzer prioritizes generating traces which
are linearizable, as explained more precisely in
Section~\ref{subsec:call-return}. It then prioritizes traces with
events making calls to the off-chain services and generates traces
which re-order calling events and its corresponding callback events
independently. If a concrete trace tested gives output the same as a
linearizable trace, it is not reported as a bug.

Section~\ref{sec:performance} explains the exponential reduction of these
optimizations on the \erc bug mentioned in
Section~\ref{sec:motivation} with experimental data.

% The task of the function collecting stage is twofold. First, from the
% bytecode it extracts all possible functions (with a simpler method
% based on pattern matching).  Second, from this set it eliminates
% functions that neither read nor write to the global variables of the
% analyzed contract -- by definition such functions cannot cause EO
% bugs. The filtrations is done with taint analysis and symbolic
% execution.

% The set of remaining functions is processed by the second stage. Each
% function is executed symbolically and from the valid executions the
% algorithm produces contexts for the function.  All pairs composed of
% functions and their contexts constitute events. They are passed to the
% third stage.

\subsection{Extracting HB Relations}
\label{subsec:extracthb}
EVM semantics do not feature programming abstractions like Java's
\code{synchronized}~\cite{jcip} or a rigorous specification of dynamic
event precedence~\cite{Bielik-al:OOPSLA15,Maiya-al:PLDI14}, which
concurrency analysis techniques targeting other languages or platforms
use. That said, smart contract developers impose an \emph{intrinsic}
ordering on events by engineering ad-hoc synchronization via a
mechanism EVM supports natively: exceptions\footnote{In Solidity,
  exceptions are raised via \fcode{throw}, \fcode{require} and
  \fcode{assert}.} and global variables. Specifically, the contract
may set values for global variables in the processing of one event,
and later check / use these before performing critical operations
under subsequent event. If the check / use of the global variable
results in an exception, the second event will not lead to valid
trace and the execution of that event will be nullified. Our
observation is that if we can detect these cases, when there is only
one valid ordering between a pair of events, we do not need to test
for any event orderings where there is an invalid order.

These ordering relations are captured formally by a happens-before
relation~\cite{Lamport:CACM78}. In all event traces, if executing an
event $e_1$ before $e_2$ leads to an exception, but executing $e_2$
before $e_1$ leads to valid (normal) execution, then we say $e_2$
\emph{happens before} $e_1$ (denoted $\hb{e_2}{e_1}$). Finally, if
neither from $\hb{e_1}{e_2}$ and $\hb{e_2}{e_1}$ holds, $e_1$ and
$e_2$ can occur in an event trace in any order and are called {\em
independent} events. The precise definition of the notion of
happens-before, as used in this paper, is below.

\begin{definition}[Happens-before]
\label{def:hb}
For a fixed contract instance $c$, we say that two events $e_1$ and
$e_2$ are in \emph{happens-before} relation $\hb{e_1}{e_2}$ with
respect to a set of valid event traces $H$ of $c$ \Iff for any trace
$h \in H$, if $h = \text{concat}(h_1, [e_2], h_2)$ and $e_1 \in h$, then
$e_1 \in h_1$. In other words, $e_1$ always precedes $e_2$ in a
trace $h \in H$, in which both $e_1$ and $e_2$ are present.
\end{definition}

Recovering the complete $\Hb$ relation (as defined above) for a
program is difficult even with a powerful symbolic analysis.  Consider
a simplified version of a real-world ~\code{Bounty} contract, shown
in Figure~\ref{fig:bounty}. The way it is implemented, an event $e_1
= \angled{\mathtt{donate}, \ldots}$ will always precede $e_2
= \angled{\mathtt{payout}, \ldots}$ in any valid execution trace,
because once $e_2$ is executed, the \code{require} clause will trigger
an exception if more events of type $e_1$ occur. This achieved by
setting the flag \code{bounty_paid} to \code{true} in line~10. There
are 2 paths in $e_2$ and 4 paths in $e_1$, and thus a total of 8
combination of event-paths to check in order to discover the complete
$\Hb$ relation between $e_1$ and $e_2$ statically. Our symbolic
analysis can encode the path's logic as SMT formula and check both the
orders of a $8$ pairs using an SMT solver for differing outputs.

\begin{figure}[t]
\begin{center}
\centering
\begin{tabular}{cc}
\begin{lstlisting}[mathescape=true,language=Solidity,basicstyle=\footnotesize\ttfamily]
contract Bounty {
  bool public bounty_paid = false;
  address public proposed_beneficiary = 0x0;
  mapping (address => bool) public has_donated;
  mapping (address => uint256) public balances;

  function payout(string _password) {
    require(!bounty_paid);
    // More requirements on _password etc
    bounty_paid = true;
    proposed_beneficiary.transfer(this.balance);
  }

  function donate() payable {
    require(!bounty_paid);
    // more requirements 
    if (!has_donated[msg.sender]) {
      has_donated[msg.sender] = true;
    } 
    balances[msg.sender] += msg.value;
  }
  // More functions...
}
\end{lstlisting}
\end{tabular}
\end{center}
\caption{\code{Bounty} contract with an intrinsic HB relation.}
\label{fig:bounty} 
\end{figure}

This approach works well for a small number of events (\eg 2 events in
this example). Extracting the full $\Hb$ relation for a larger set of
events quickly leads to a combinatorial explosion of path orders to
search. To address this challenge, our approach infers a ``weaker''
form of HB-relations, which operates on pairs of events, considering
traces with only two events. The following definition makes precise
our design choice, and enables a direct implementation strategy.

\begin{definition}[Weak happens-before]
\label{def:whb}
For a set $E$ of events of a contract instance
$c = \angled{\Id, M, \cstate}$ and a blockchain state $\bstate$, such
that $\bstate[\Id] = \cstate$, we say that two events $e_1, e_2 \in E$
are in a weak happens-before relation ($\whb{e_1}{e_2}$) \Iff (a)
$\cstate \tstep{[e_1, e_2]} \cstate'$ for some contract state
$\cstate'$, and (b)
%
%$\cstate \estep{e_2} \cstate''$ for some $\cstate''$, and (c)
%
$\cstate \tstep{[e_2, e_1]} \exn$.
\end{definition}

%  In EVM, exceptions allow one to specify a dynamic check, prohibiting
%  execution of a certain event at run-time in a given contract
%  state. Therefore, exceptions are commonly used to implement a
%  functionality, similar to an atomic \emph{compare-and-swap} (CAS)
%  primitive in concurrent architectures~\cite{Herlihy-Shavit:08},
%  inducing a happens-before ordering.
% 

The implementation strategy for extracting weak happens-before (WHB)
is straight-forward. We execute the two differing orderings of each
pair of events in a given trace. If we observe that one order leads
to exception, and other does not, we inductively learn this
relationship. We can also identify which functions are independent as
per the natural extension of the definition from HB to WHB.

Notice that WHB implies regular HB for traces of two events, but HB
for longer traces can include fewer pairs due to state changes made by
other events in a trace.  Using WHB leads to an under-approximation
of the full $\Hb$ relation, as it may render more pairs of
events \emph{non-independent}. This could introduce more false
negatives (\ie, will miss some EO-bugs), but at the advantage of an
exponential reduction in trace combinations to test.

An important final optimization helps the symbolic analysis of event
pairs. \toolname builds weak \HB-relation for events pairs that share
at least one global variable and at least of the functions write to
the shared variable. Only such functions when re-ordered may lead to
different values of global storage variables and thus their
corresponding events may be in weak \HB-relations. \tool analyzes the
symbolic write sets of program paths, and uses this information to
guide which events to consider during $\Hb$ analysis.

% \begin{comment} Further we present a computer tool called \toolname
% that helps to detect execution-ordering bugs in Ethereum smart
% contracts. To analyze a contract, \toolname takes as input only
% contract's bytecode and outputs a list of pairs of valid event
% histories that constitute event-ordering bugs.

% To address the problem of feasibility and remain practical while
% scanning a vast amount of potential histories and their orderings,
% the tool first uses static analysis to significantly reduce the
% search space, and then, dynamic analysis to test the potential
% candidates. The former is accomplished with symbolic execution,
% while later with independent implementation of the Ethereum Virtual
% Machine.

% \begin{definition}[Significant function]
% \end{definition}

% \begin{definition}[Potential HB pair of functions]
% \end{definition}

% \end{comment}

\subsection{Optimizing for Linearizability Checking}
\label{subsec:call-return}
This optimization is based on the observation that certain orderings
of events are well-paired by EVM semantics. When contract execution
adheres to these prescribed paired orderings, the results are what the
developer likely intended (and must not be flagged). In fact, all
orderings which lead to one of these intended outputs do not highlight
an unintentional behavior, and need not to be reported. This
optimization reduces the witness pairs finally reported to the human
analyst, suppressing results that are likely benign.

\begin{figure}[t]
\centering
\includegraphics[width=0.5\textwidth]{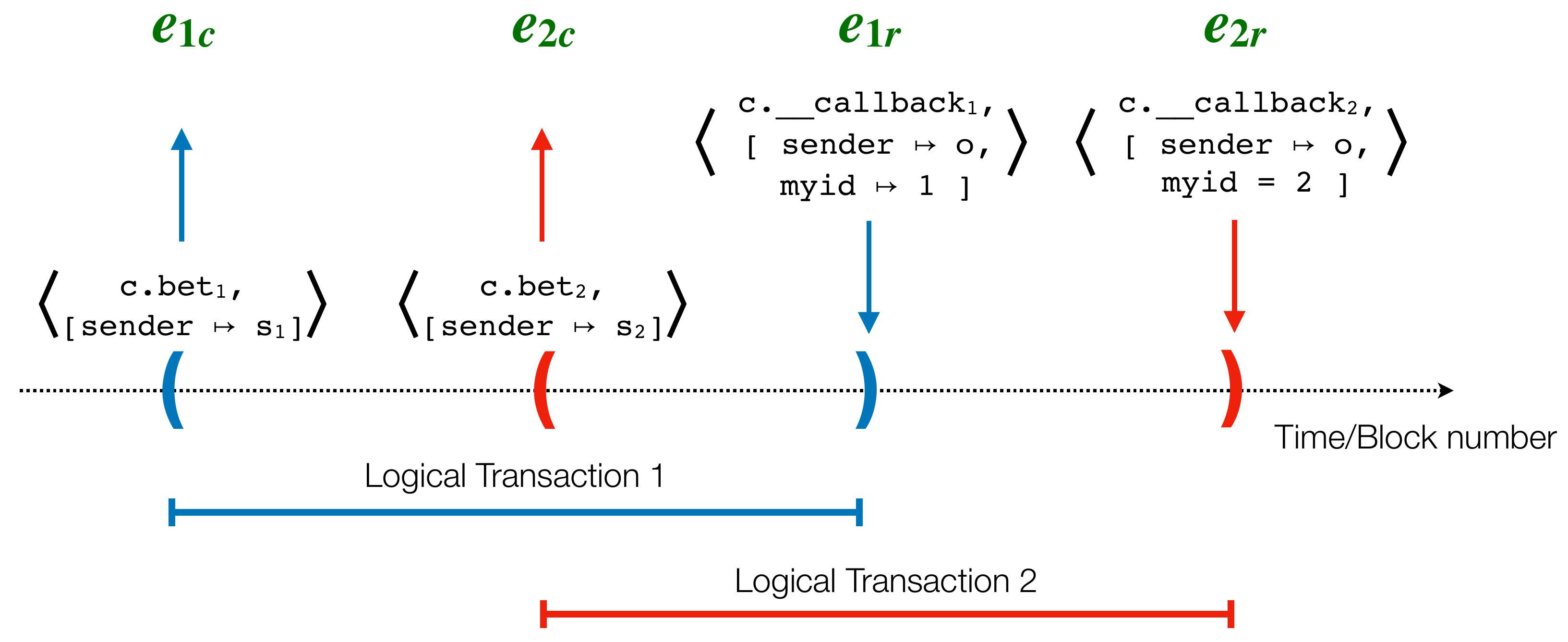}
\caption{A call/return event trace in \code{Casino} contract.}
\label{fig:atomic}
\end{figure}

At present, \tool uses such optimization for asynchronous call to
\Oraclize.  For interacting with off-chain resources (\eg, stock
exchanges, random number generators), a number of Ethereum contracts
use the popular service \Oraclize~\cite{oraclize}.  \Oraclize handles
the query \emph{offline}, by retrieving the data from the required
URLs, and later sends it back to the asking contract by calling its
function \code{__callback} in a separate Ethereum
transaction.\footnote{A contract calling \Oraclize must have a
  function called \texttt{\_\_callback}.}  Hence, a call to \Oraclize
would ``return'' via an asynchronous callback.

Here, for each \texttt{Oraclize} call, there is matching return or
callback. Programmers are expected to match returns to the \Oraclize
calls, and process accordingly, handling multiple users.  The notion
that captures intent (or natural program reasoning) is
linearizability~\cite{Herlihy-Wing:TOPLAS90}: each pair of matching
call-returns must appear to execute atomically. That is, if two users
issue events that trigger \Oraclize calls, then either the first
user's call-returns are executed before the second user's, or the
other way around; if their call-return strictly interleave
out-of-order, it leads to an unintended (or unnatural) behavior.  This
is evident in the \code{Casino} example (Figure~\ref{fig:casino}).

\tool in its fuzzing step recognizes \Oraclize calls and asynchronous
returns. It first checks all linearizable traces, where matching
call-return pairs execute atomically. It memoizes the outputs of these
linearizable traces as ``canonical'' outputs. Then, it generates
other possible tracees during its testing and reports a witness pair
only if the outputs are not equal to all of the canonical outputs.  If
a trace is found which does not match any linearized event trace,
\tool flags the contract as having a EO bug, and reports on the
closest linearizable trace in the witness pair. One example of such
non-linearizable trace for the \code{Casino} contract is presented
in Figure~\ref{fig:atomic}.
%
% Given this definition, when an event history corresponds to a buggy
% execution in this model? The answer is: when a logical transaction can
% be observed to be non-atomic, \ie, a client (miner) can construct a
% scheduling of events that would result in a contract state, which
% \emph{could not} take place if every call-event was immediately
% followed by the corresponding return-event.
% %
% As the reader might notice, this idea is not new---it is merely a
% rephrasing of the classical definition of
% \emph{linearizability}, a golden standard
% in specification of concurrent data structures. What is new, is the
% application to smart contracts, made possible thanks to our event
% model. The formal definition is, therefore, as follows:
% %
%
For completeness, we provide a definition of linearizable traces,
which precisely capture this notion.

\begin{definition}[Linearizable event traces]
\label{def:lin}
  A valid event trace $h = [e_1, \ldots, e_n]$ of a contract
  instance $c = \angled{\Id, M, \cstate_0}$ is linearizable \Iff 
  a state $\cstate_n$, such that $\cstate_0 \tstep{h} \cstate_n$, can
  be obtained by executing a trace $h'$, which is a permutation of
  events in $h$, such that
  \begin{enumerate}
  \item it preserves the order of call/return events from $h$;
  \item non-overlapping logical transactions in $h$ appear in the same
    order, as they appear in $h$.
  \end{enumerate}  
\end{definition}

%\noindent

% We now formally characterize this class of erroneous contract
% behaviors and extend \toolname for detecting them.
% %demonstrated by the \code{Casino} example from Figure~\ref{fig:casino}.
% %
% To do so, let us refine the definition of contract events for a
% specific implementation pattern, which encodes asynchronous
% callback-based programming and is typical for interaction with
% external oracles.
% 
% % \footnote{In all definitions for the rest of this
% %   section, we consider a fixed contract instance of interest, as well
% %   as the initial blockchain state, unless mentioned otherwise.}
% 
% \begin{definition}[Call/Return events]
% \label{def:call-return}
%   For a given valid history of events $h = [e_1, \ldots, e_n]$ of a
%   contract instance $c = \angled{\Id, \cstate_0}$, two events, $e_i$
%   and $e_j$ ($1 \leq i \le j \leq n$), encode a \emph{call/return
%   pair} 
%   %
%  \Iff there exists a contract $\Id'$, such that
%   %
%  \begin{enumerate}
%  \item There exists a value $v$, such that for
%    $\cstate_{i-1} \estep{e_i} \cstate_i$, it holds that
%    $v \in \cstate_i$ and $\Id' \in \cstate_i$, and
%    %
%  \item for $e_j = \angled{\fn, m}$, $m.\Sender = \Id'$ and
%    $v \in m.\Data$.
%  \end{enumerate}
% \end{definition}
%
\noindent
%
% In both clauses Definition~\ref{def:call-return} we abuse the
% $\in$-notation to denote, in (1), that both value $v$ and contract
% address $\Id'$ are recorded in the resulting state $\cstate_i$ of
% $\Id$, and, in (2), that the message $m$ of $e_j$ comes from $\Id'$
% and contains $v$ in its payload.
% %
% In the definition, an account $\Id'$ plays a role of a ``concurrent
% environment'', to which the contract $\Id$ transfers control upon
% executing $e_i$ (and, conversely, from which it gets control via
% $e_j$); the value $v$ serves as a ``call identifier'' and is typically
% a hash, allowing the contract implementation \emph{match} an outgoing
% call with an incoming return---precisely what is done via
% \code{oid}/\code{myid} in the implementation in
% Figure~\ref{fig:casino}.

% Together, pairs of call/return events correspond to ``logical
% transactions'', \ie, contract executions spanning two blockchain
% transactions. An example of a history
% $h = [e_{1c}, e_{2c}, e_{1r}, e_{2r}]$ with two overlapping logical
% transactions and the corresponding four call/return events for
% \code{Casino} is shown in Figure~\ref{fig:atomic}.
%
{

\section{Implementation}
\label{sec:implementation}

\toolname is implemented in about 6,000 lines of Python. 
To process a contract, the tool requires as its input either a
contract's bytecode or its Ethereum address. If the latter is given,
\toolname assumes the contract is deployed on the main Ethereum blockchain and gets its
bytecode and current global storage from the blockchain, and stores it locally.
Otherwise, it assumes the contract is not deployed, and thus starts
with empty local storage.  An optional input to \toolname is
contract's Solidity source code. If provided, it improves the
readability of the produced reports, allowing \toolname to use
function names instead of EVM signatures.\footnote{A function
  signature is an 8-digit hexadecimal number that uniquely identifies
  functions when compiled from Solidity source.} The tool reports all
pairs of traces that lead to EO bugs.

\subsection{Preprocessing}
\label{sec:extr-funct-events}

From the contract's bytecode, \toolname first collects all the
function signatures in it. This is done via a heuristic search that
inspects the beginning of the bytecode, where all signatures are
defined~\cite{Gavin-al:yellow-paper}, and matches it against
predefined patterns.
Alternatively, if Solidity source code is provided, then the tool
directly extracts the signatures from the source by compiling the code
with Solidity compiler \texttt{solc} with the \texttt{--hashes} switch
on. 
%To the list of signatures the tool manually add the fallback function.\footnote{As a signature of the fallback, \toolname uses is either \texttt{0x11111111} or \texttt{0x22222222}.}

\begin{comment}

\subsection{Extracting Functions}
\label{sec:extr-funct-events}

As its very first task, \toolname, when provided with a contract
bytecode, first collects all the function signatures in it. This is
done via a heuristic search that inspects the beginning of the
bytecode, where all signatures are
defined~\cite{Gavin-al:yellow-paper}, and matches it against
predefined patterns.
%
Alternatively, if Solidity source code is provided, then the tool
directly extracts the signatures from the source by compiling the code
with Solidity compiler \texttt{solc} with the \texttt{--hashes} switch
on. In the later phases of the \toolname pipeline, 
%
\todo{Refer to specific part of Figure~\ref{fig:tool}.}
%
function signatures are
used as unique identifiers of the function components in events. To
the list of signatures the tool manually add the fallback
function.\footnote{As a signature of the fallback, \toolname uses is
  either \texttt{0x11111111} or \texttt{0x22222222}.}

\end{comment}

\subsection{Dynamic Symbolic Execution Engine}
\label{sec:symbolic-execution}

\toolname relies on standard DSE, improving the open-source
implementation of a recent work~\cite{Nikolic-al:Maian}.  All inputs
to the entry points of contract bytecode are marked symbolic
initially. The symbolic execution starts from the first instruction of
a contract's bytecode and interprets sequentially the instructions
following the EVM specification~\cite{Gavin-al:yellow-paper}, with
access to running stack, global storage, and local memory. The DSE
engine keeps two memory maps: a symbolic map and concrete local map. The local
map is initialized with the concrete blockchain state given as input
to \toolname.  All global variables of the contract are concretely
initialized from the local copy of the global storage. The DSE engine
gathers symbolic values, branch, and path constraints in the standard
way. We check satisfiability of symbolic constraints using
Z3~\cite{deMoura-Bjorner:TACAS08}, which can handle operations on
numeric and bit-vector domains. Our present implementation supports
symbolic analysis of $90\%$ of all EVM opcodes. The unhandled
instructions preclude analysis of a small number of paths in $6\%$ of
the contracts we analyzed.

The analysis aims to over-approximate values in its symbolic path
constraints as a default strategy. For instance, when checking with
path feasibility with an SMT solver, if the solver times out, we
assume that the path is feasible. We prune away paths only if the SMT
solver returns \texttt{UNSAT}. We over-approximate the set of pointer
values during the analysis for symbolic memory by not concretizing it
eagerly.  Similarly, the DSE engine marks return values of
\texttt{CALL} instructions (which invokes functions from other
contracts) as symbolic. During the DSE analysis (and only during this
analysis), we do not concretely execute the externally called
contract, but the symbolic return value over-approximates the
behavior. External calls cannot modify the caller's contract state,
hence marking only the return value as symbolic is sufficient.

While our symbolic analysis largely over-approximates, it concretizes
in a small number cases which help improving path coverage. First,
when the value of the symbolic variable is needed for checking path
feasibility, we lazily concretize it from the concrete to drive
execution down that branch.  Second, the DSE engine concretizes
lookups for key-value stores (\texttt{hashmap}/\texttt{hashtable}
types), which are called ``mappings'' in Solidity~\cite{solidity} and
implemented using \texttt{SHA3} as the hash function. Our baseline
system~\cite{Nikolic-al:Maian}, did not have support for this, and
this implementation detail is crucial for finding EO bugs in many
contracts, including the \erc example from Figure~\ref{fig:ERC20}.  If
the lookup key $v$ is symbolic (\eg, it came as a contract call
input), our DSE engine concretizes it to the set of values assigned to
it in concrete executions observed during DSE. In addition, the first time
a symbolic $v$ is accessed in a write, we assign it a new random
concrete value so that the concrete value set is never empty.

As an example, in Solidity, for a mapping \code{s} of type
\code{address => int}, the value of $s[v]$ is located in the global
storage at address $\mathtt{SHA3}(v ~@~ \Index(s))$, where $@$
denotes concatenation of words and $\Index(s)$ returns the identifier
of a variable \texttt{s} in the contract.\footnote{Each global
  variable has an identifier that depends on the order of how the
  global variables have been defined in the contract.}  For each
$\Index(s)$, the tool maintains a list of all concrete keys $v$
\emph{seen} during the previous executions. When having to execute
$\mathtt{SHA3}(v ~@~ \Index(s))$ with a symbolic $v$, it retrieves the
list $\mathit{vs}$ of all seen concrete values of $v$ for $\Index(s)$,
branches the execution on all such values, and executes concretely
$\mathtt{SHA3}(v ~@~ \Index(s))$ for every $v \in \mathit{vs}$.  In
addition, whenever required, \tool generates a random value for $v$, adds it to $\Index(s)$. 

% Mappings store their values at random addresses (computed
% with $\mathtt{SHA3}$), but they do not provide a way of finding which
% are their non-empty keys. During symbolic execution, often we need to
% check the value of a mapping at keys\footnote{The key is often an
%   address (as for the mapping $s$ in the example above).} given as
% symbolic variables. The heuristic will help to handle this case.

\paragraph{Side-Effects \& HB Relations}
During the DSE analysis, for each collected function, \toolname
performs conservatively records locations of all global variables of
the contract to which the function reads or writes.  This analysis
helps optimize away functions which do not update global state, as
discussed previously. It aid subsequent extraction of $\Hb$ relations
even further as explained previously. The functions that
do not read nor write to any shared global variables need not be
considered  for $\Hb$ analysis. 

The $\Hb$ analysis of symbolic paths is straight-forward. Paths
leading to exceptions are identified by special
instructions.\footnote{These include instructions that do not have a
  valid bytecode, or jumps to unlabelled code indexes (for instance, $0000$).}  For
a pair of symbolic events and corresponding path constraints, we check
for path feasibility. The SMT solver returns concrete events that
validate if a particular order triggers an exception.

\subsection{Fuzzing Event Traces}
\label{sec:fuzzing}

Given the set $E$ of all events produced previously, the fuzzer engine
creates all possible subsets of $E$ up to a certain size.\footnote{The
  maximal size is parametrized, thus it can be increased or reduced;
  we tried sizes from 2 to 6 in our experiments.}
For each subset, all possible traces, which obey the previously
discovered HB relation, are generated. 
Each such trace is executed on a customized EVM instance, which in comparison
to the original EVM (of original Ethereum client) is orders of magnitudes
faster, as it runs without performing the proof-of-work mining, and does not
participate in a message exchange with other nodes on the network.
The customized EVM, however, can read from the original Ethereum
blockchain by sending queries to Ethereum \texttt{geth} client running on a
local machine and participating in the Ethereum network. Hence, if \toolname
processes a contract deployed on the main Ethereum blockchain, the customized EVM initially
reads its actual global state, stores it locally, and uses this copy for all sequential reads and writes of the global storage.

After executing each trace, the global storage and the balance of
the contract are saved. Once all traces for a particular subset
have been executed, the tool finds and outputs a list of pairs of
EO-bug traces, \ie, pairs of traces that result either in
different global storage or in different balances.
%
\begin{comment}
\todo{Why this is not done immediately for each history?}
\inik{Done what, compare to itself? 
\\
We can always take a history, produce all permutations and compare. But then, the complexity will be much higher as histories will be repeating. Instead, we create all possible permutations of certain subset of events and compare if any two such permutations can lead to different storage. 
}
\end{comment}
%
The tool then performs minimization by creating a set of pairs of \emph{minimal
  traces} of function calls, reproducing the found EO-bugs. 
The minimization is done by implementing a simple shrinking strategy:
same events are removed from the pair of buggy traces, one by one, while
checking whether a ``smaller'' pair of traces still constitute an EO-bug.
In our experience, the size of the set of minimal traces is
significantly smaller 
%\todo{By how much?} 
% at this stage is early to talk about concrete numbers
than the length of the full
list of buggy traces, so having the minimization procedure provides
a better user experience and allows for faster confirmation of
witnesses by a human.

To further reduce the number of benign violation of EO, \toolname implements
an additional simple heuristic. \tool gives priority to checking all
permutations with distinct events. Permutations that reorder events with the
same entry point but different input values are checked after all others. The
rationale behind this heuristic comes from the fact that a large number of
functions change the global output state by writing a value that depends on
some input parameter. Thus, calling such functions with different inputs will
always result in different global storage of the contract.  This is often
intended behavior, rather than a bug.

% A trace and its reordered
% version are tested if they result in the same global storage or
% balance only if in each of the traces, the sequences of events to
% the \emph{same} function coincide. In other words, the reordered
% trace can be produced from the original only by reordering different
% functions, but not by reordering the \emph{inputs} to the \emph{same
%   function}. 
%
\begin{comment}
\todo{I'm confused by this: how reordering functions is different from
  reordering inputs? What if I have events $\angled{f, m_1}$ and
  $\angled{f, m_2}$ and try them in a different order --- does it
  count as reordering functions or inputs?}
  
Does this really matter?   
  
\end{comment}
%

%
\begin{comment}
\todo{Still, I got no answer---what happened to the discussion on
  disjoint footprints?}

What is a disjoint footprint? 
  
\end{comment}

\begin{figure*}[t]
%\setlength{\belowcaptionskip}{-5pt}
      % \centering
      % \begin{minipage}{0.3\linewidth}
  \begin{subfigure}[t]{0.33\textwidth}
      \includegraphics[width=\linewidth]{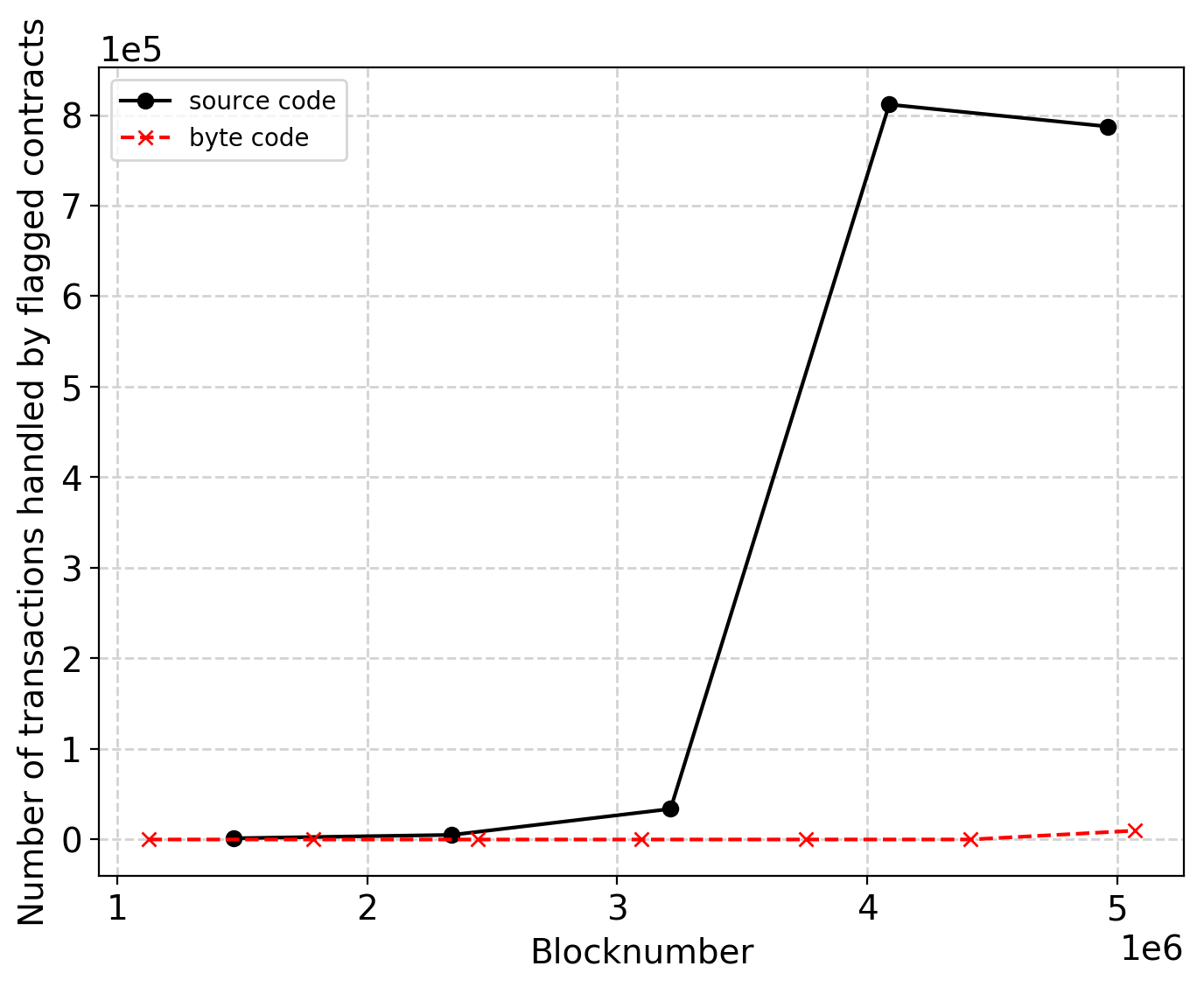}
      \caption{Activity of contracts.}
      \label{fig:activity}
  \end{subfigure}
% \end{minipage}
  \hspace{0.005\linewidth}
% \begin{minipage}{0.3\linewidth}
  \begin{subfigure}[t]{0.33\textwidth}
      \includegraphics[width=\linewidth]{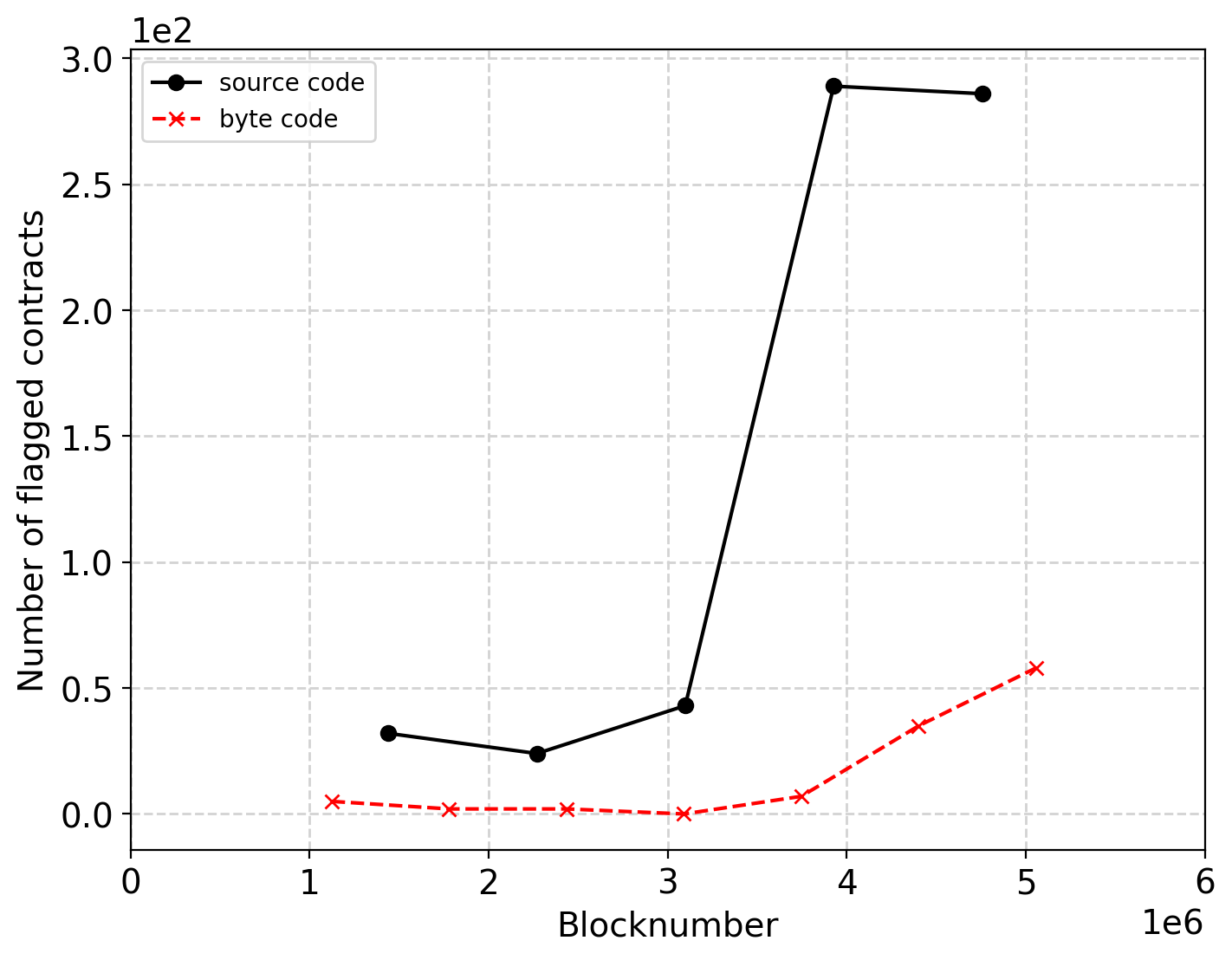}
      \caption{Recency of contracts.}
      \label{fig:recency}
  \end{subfigure}
% \end{minipage}
  \hspace{0.005\linewidth}
% \begin{minipage}{0.3\linewidth}
  \begin{subfigure}[t]{0.32\textwidth}
    \includegraphics[width=\linewidth]{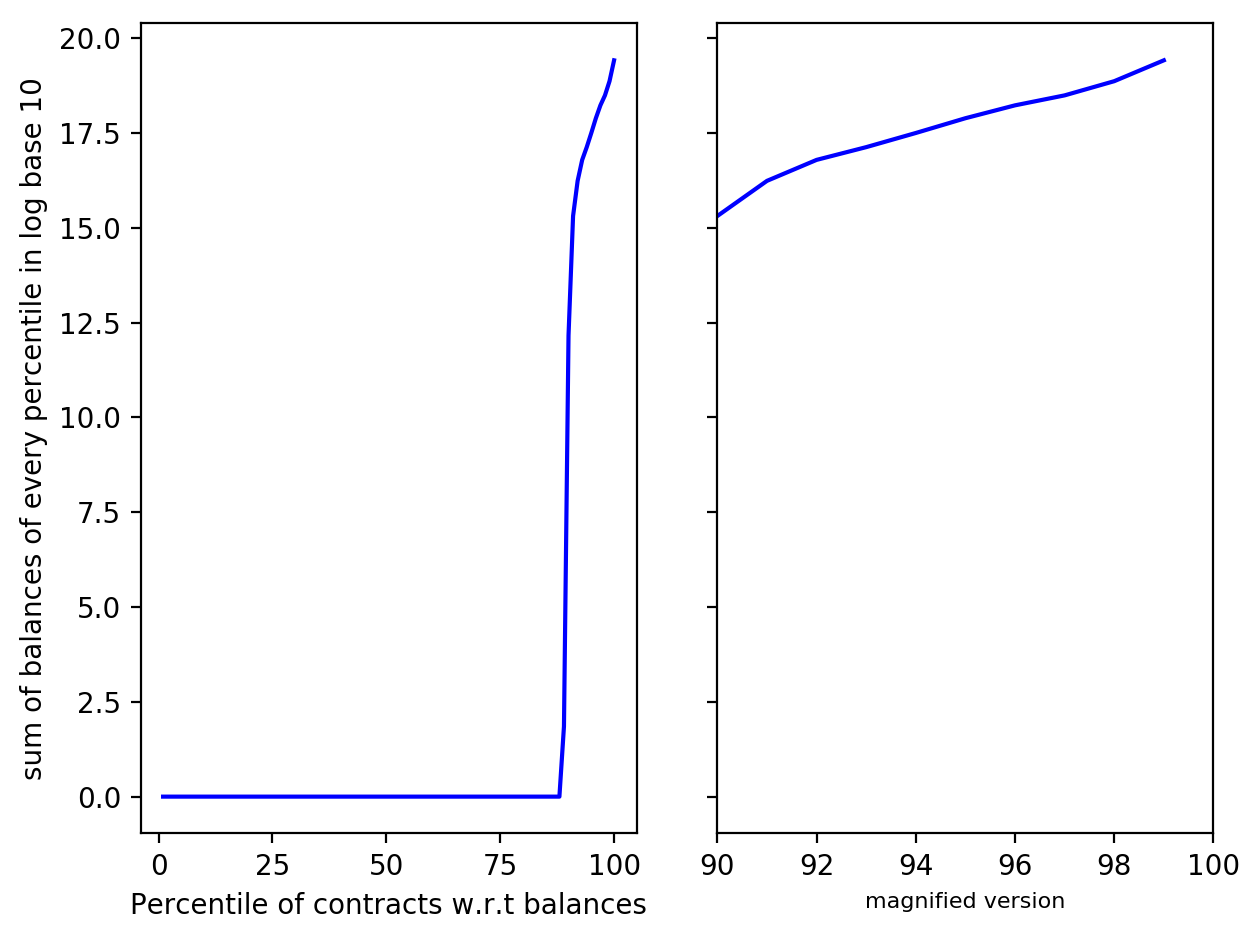}  
    \caption{Held balances.}
    \label{fig:baldist}
  \end{subfigure}
      % \end{minipage}
  \caption{Statistics for the analyzed contracts.
    The histogram plot~\ref{fig:activity} provides the total transactions
    handled by the flagged contracts with respect to the block
    number. From block number $3,000,000$ to $5,000,000$, contracts
    with source code have seen tremendous increase in the amount of
    transactions, whereas for contract with only bytecode this number
    has remained almost constant. 
    Similarly, histogram plot \ref{fig:recency} shows the number of contracts created with respect
    to block number. As evident, most of them have been created
    recently, \ie, after block number $4,000,000$.  
    Finally, the plot \ref{fig:baldist} shows the ``accumulated'' sum of balances of
    contracts. The highest $1\%$ contracts according to their balance
    make up for more than $99\%$ of the ether in the flagged
    contracts.
  }
  \label{fig:conchar}
\end{figure*}

\subsection{Checking Linearizability of Event Traces}
\label{sec:line-up}

The \oraclize component of \toolname is triggered on \Oraclize-using
contracts, It tailored for detecting a specific kind of EO-bugs,
namely linearizability violations (see Definition~\ref{def:lin})
during fuzzing.
For each contract, the \oraclize considers only two functions: one
that sends a query to \Oraclize and another that receives the
``matching'' return from \Oraclize. The former can be arbitrary, but
the latter is always declared as a function with the signature
\code{__callback(bytes32 _queryId, string _result,... )},
where \code{_queryId} is a \emph{unique} identifier of the query to
\Oraclize.

Even though the calling function can be arbitrary, the \oraclize finds
the matching call/return pair by inspecting the blockchain in the
following way. Each call to \Oraclize writes a unique \code{_queryId}
to the blockchain called \emph{a log},\footnote{In Solidity, these are
  called Events, not to be confused with our contract events from
  Definition~\ref{def:event}.}
while a client contract also stores this value in its persistent state
(to use it to identify a matching call from \Oraclize).
Therefore, for each \code{_queryId}, the matching reply from \Oraclize
can be found by inspecting the incoming calls of the analyzed
contract: a call that originates from \Oraclize and invokes the
\code{__callback} function with the required value of \code{_queryId}
as first argument is the match.
Rather than producing new contexts, the \oraclize retrieves them from
the blockchain when scanning for all values of the \code{_queryId}
parameter.
%
%
\begin{comment}
Producing useful contexts for \code{__callback}-events is impossible
without manually inspecting a contract, because \todo{can we have a
  simple explanation here?}

\end{comment}

% \footnote{The required queried URL for an
%   arbitrary contract cannot be determined, thus the reply from
%   \Oraclize may not have the format required for the contract to
%   function properly.}

\section{Evaluation}
\label{sec:evaluation}
We evaluate \toolname to measure how many contracts are flagged as
having EO bugs, \ie, how many show differing outputs under a different
ordering of events. Further, we measure how much effort the analyst
has to spend per contract to analyze the flagged traces. We asses
how EO bugs compare to existing definitions of transaction re-ordering
bugs checked by the existing \oyente tool.  Lastly, we present real
case studies highlighting efficiency of \toolname. We highlight only important findings in this section and provide more examples and case 
studies in Appendix~\ref{appendix:casestudies}. We urge the readers 
to refer to the Appendix to gain better insights on our results.

% %
% 
% Due to space constraints we detail only the important findings in this
% section, and provide graphs, examples and similar relevant data in
%  Appendix~\ref{sect:appendix:oraclize}. % -- we urge the interested readers to refer to the Appendix to gain better insights of our results. %, wherever necessary.
% 
% % We explain how our tool compares to \oyente, which flags a related
% % class of bugs called transaction order dependency (TOD) bugs. We
% % point out that definitions of TOD are subtly different from our
% % characterization of EO bugs, leading to substantially different
% % results and false positive rates.

\paragraph{Evaluation subjects}
Around $1\%$ of the smart contracts deployed on Ethereum have source
code.  \toolname can directly operate on EVM bytecode and does not
require source. To test the effectiveness of the tool, we select
$5,000$ contracts for which Solidity source code\footnote{We obtain
  contract source codes from the popular website
  Etherscan~\cite{etherscanSourceCodes},} is available and another
$5,000$ contracts randomly chosen from the Ethereum blockchain (for
which source is unavailable). 

\toolname takes all these contracts in bytecode form
and analyzes them for \syn violations. On the other hand, for analysis
of \lin violations, we find $1,152$ unique contracts on the Ethereum
blockchain to which \Oraclize callbacks have occurred. We
filter out contracts with less than two \Oraclize queries, and
are left with $423$ unique contracts out of which $154$ have Solidity
source code available. We analyze these $423$ contracts for \lin
violations.

\paragraph{Experimental Setup}
We run all our experiments on a Linux server, with 64GB RAM and 40
CPUs Intel(R) Xeon(R) E5-2680 v2@2.80GHz. We process the contracts in
parallel on $30$ cores, with one dedicated core per contract. We set a
timeout of $150$ minutes per contract. \toolname is configured to
output $3$ pairs of events for each pair of functions in the HB
relation from its an analysis component; the timeout per pair of
functions in the $\HB$ relation is $2$ minutes. For each function with no
event generated in the $\HB$ creation phase, \toolname either generates
one event or times out in $1$ minute.  These events are used for the
fuzzing step in \toolname.  The outputs of the \toolname are a set of
pairs of event traces, flagged as having EO bugs. We plan to
release a public artifact to enable direct reproducibility of the
reported experiments.

\subsection{Efficacy of \toolname}

\tool flags a total of $789$ ($7.89\%)$ \syn and $47$ ($11\%$) \lin violations
in the analyzed smart contracts, holding hundreds of millions of dollars worth
of Ether over their lifetime. Figure~\ref{fig:conchar} provides statistics on
the transaction volumes, recency and  ether held by these contracts.

% \textbf{Figure X provides statistics on the Ether held, recency of
% these contracts and ...}

\paragraph{\Syn violations} 
Out of the $10,000$ analyzed contracts, $789$ are flagged as having \syn bugs.
This shows that a large majority of contracts do not have differing outputs
upon the tested orders.  Among the $789$ contracts, $674$ have source code,
which we manually analyze, and $115$ are from the set of contracts without
source.
The maximum Ether held by the contracts over their lifetime sums up to
$703,676$, which is approximately $142.5$ million
USD.\footnote{Calculated at 203 USD per one Ether.}  A large fraction
of this amount belongs to contracts that are based on the \erc
standard, as the example presented in Section~\ref{sec:motivation}.

At present, $785$ out of $789$ contracts are live and currently hold
$307$ Ether ($62,321$ USD equivalent); for the distribution of Ether
in these contracts).  The flagged contracts are transaction intensive
as shown in Figure~\ref{fig:activity} and the volume of processed
transactions over their lifetime amounts to $1,649,192$ transactions
on the public Ethereum blockchain.

Recent contracts percentage-wise have fewer bugs, presumably due to
the rise of automatic analysis tools and manual audit firms.  However,
shear increase in the number of deployed contracts results in larger
number of flagged contracts in recent contracts. This trend can be
seen in Figure~\ref{fig:recency}.

% 
% \begin{comment}
% Figure~\ref{fig:baldist} shows the distribution of Ether held in these
% contracts. The maximum Ether held by the contracts over their lifetime
% sums up to $703,676$, which is approximately $551$ million
% USD\footnote{Calculated at 784 USD per one Ether.} as of this
% writing. A large fraction of these belong to contracts that are based
% on the \texttt{ERC-20} contract, one example of which is presented in
% Section~\ref{sec:motivation}. Figure~\ref{fig:activity} shows %the
% usage of these contracts by the volume of transactions processed by
% these contracts over their lifetime. In short, the $789$ flagged
% contracts have a total of $1,649,192$ transactions on the Ethereum
% public blockchain.
% 
% One may expect more recent contracts to have fewer
% number of bugs because of the rise of automatic analysis tools and manual audit firms. 
% % \inik{I cannot edit the paragraph below without the actual figure.} 
% Figure~\ref{fig:recency} shows the recency of the tested and vulnerable contracts, by showing
% the point at which they are created. $785$ out of $789$ contracts are
% live at present and presently hold $307$ Ether ($240,688$ USD
% equivalent). It can be seen from Figure~\ref{fig:recency} that the number of flagged
% contracts increases with recency (\ie,  with the block height),
% showing that many recent contracts are buggy. 
% 
% \end{comment}

% \vspace{-5pt}
\begin{figure}[t]
{
\setlength{\belowcaptionskip}{-15pt}
\includegraphics[width=\linewidth]{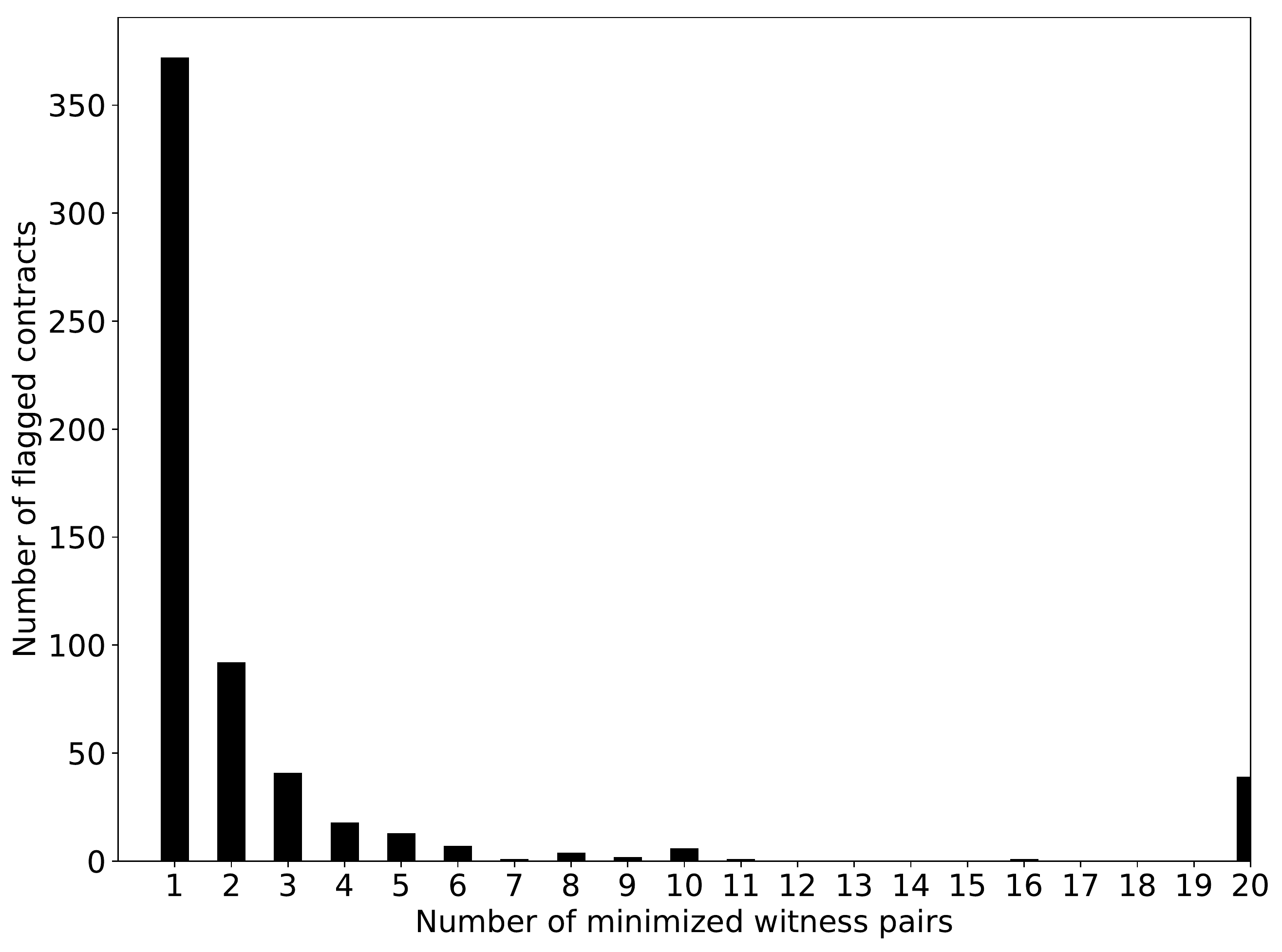}  
\caption{Number of minimal traces in flagged contracts. }
\label{fig:trace_sol} 
}
\end{figure}
% \vspace{-5pt}

\paragraph{\Lin violations}
Out of the $423$ analyzed contracts, the tool flags $47$ contracts for
\lin violations. In the past, the flagged contracts have held up to
$5,067$ Ether ($1$ million USD equivalent). Currently, $28$ of them
are live and hold $862$ Ether. The contracts are  transaction
intensive as well, as they have handled $55,946$ transactions in their lifetime.

%
% Figure~\ref{figureL} shows the volume  of transactions processed by these
% contracts over their lifetime. 
% Figure~\ref{figureM}  shows the recency of these contracts. Note,

% \textbf{Repeat the above
%   analysis for these guys}.

% \vspace{-5pt}
% %

\subsection{Manual Analysis Effort}
\toolname produces as output a set of pairs of traces exhibiting EO
bugs. As shown in Figure~\ref{fig:trace_sol}, for flagged contracts
with \syn bugs the tool produces only a few minimized witness pairs,
\ie, on average two, and only a single pair for the majority of the
flagged contracts. Similarly, the contracts with \lin bugs have only
one witness pair for all vulnerable contracts.  These results
highlight that the post-hoc analysis manual effort is minimal and
generally requires inspection of only a handful of pairs of traces
by a human auditor.

To confirm the simplicity of the post-hoc analysis and to determine
whether the EO violations are intentional (benign) or buggy, we
manually analyze the flagged cases. We take a $100$ randomly chosen
contracts flagged for \syn violations and all $27$ contracts (with source code) flagged
for \lin violations.  For each of the inspected contracts, we need
only a few minutes to confirm or disprove whether the EO violation is
true bug or likely benign.

\begin{figure}[t]
{
\setlength{\belowcaptionskip}{-15pt}
{
%\centering
\begin{lstlisting}[mathescape=true,language=Solidity,basicstyle=\footnotesize\ttfamily,	xleftmargin=2em,]
contract ERC721  {
  function addPermission(address _addr) public
    onlyOwner{
      allowPermission[_addr] = true;
  }  
  function removePermission(address _addr) public
    onlyOwner{
      allowPermission[_addr] = false;
  }
  ...
}
\end{lstlisting}
\caption{False positive: fragment of an \scode{ERC721} contract.}
\label{fig:perm}
  }}
\end{figure}

\paragraph{\Syn Violations} 
For the 100 contracts we analyzed, $52\%$ are closer to the examples
presented in Section $2$ and we adjudged them to be true bugs. Apart
from the example presented in Section~\ref{sec:motivation}, we find
several other examples of subtle true EO violations.
% \footnote{Two other true positive examples are given in the anonymized
%   supplementary material:
%   \url{https://www.dropbox.com/s/yupz3qjlebrbyzs/ethracer_case_studies.pdf?dl=0}.}
%
Two additional examples of such contracts named
\code{Contest} and \code{Escrow}, are presented in
Appendix~\ref{appendix:casestudies}. 
% %
% supplementary material.

%
We deemed that $48\%$ of the $100$ contracts we analyzed were
violations of the EO definition, but likely benign. These contracts
were straight-forward to analyze, as they have a small set of global
variables shared between functions of the traces which take different
values upon re-ordering. For instance, consider the ERC721 contract
shown in Figure~\ref{fig:perm}. 
Functions \code{addPermission} and \code{removePermission} are called
by the contract owner to update the common variable
\code{allowPermission}.

\toolname indicates an EO bug with both orders of a pair of events:
(\code{addPermission}, \code{removePermission}). We adjudge such race
conditions to be intentional and benign.

% \textbf{A couple of examples would be very useful to
% explain this point}.

\paragraph{\Lin Violations}
For these violations, among the $47$ flagged contracts, $27$ have
Solidity source code, which we manually analyze. $23$ of these
contracts have true and subtle EO violations, which follow two
predominant patterns:
\begin{itemize}
\item \textbf{Unprocessed \code{_queryId}:} The first class occurs
  in contracts that do not process \code{_queryId} in their
  \code{__callback} function. Rather, the contracts assume
  synchronous nature of \Oraclize, \ie, assume that each query to
  \Oraclize is immediately followed by a reply from \Oraclize. The
  contract \code{BlockKing}, mentioned in a few previous works~\cite{Sergey-Hobor:WTSC17}, is
  an example of such contract. We have identified one more contract,
  called \code{Gamble}, that suffers from a similar mistake.

\item \textbf{Improper check on Ether:} This second class of bugs is
  more subtle and has not been identified in prior works. It occurs in
  contracts similar to our \code{Casino} example from
  Section~\ref{sec:motivation}. Upon receipt of a bet from a player,
  the contract checks if it has a sufficient amount Ether for payout
  in case \Oraclize returns value winning for the user, however, it
  does not keep track of the total amount of Ether required for payout
  of all awaiting players. Therefore, when several players have open
  games, and all of them win, the casino may not be able to pay all of
  them. On the other hand, when each of the players participates and
  gets the result immediately (\ie, when the case has been
  linearized), if the casino cannot pay the player, it does not accept
  the bet, hence the balances of the players will be different. We
  found 21 contracts that have this type of mistake.

\end{itemize}

\noindent
Out of $27$ contracts flagged for \lin violations, $4$ contracts have
EO violations which seem to be benign (or intended logic) in
contracts. Specifically, they have two or more traces that differ
in output dependent on the timestamp of the mined block. 

% A list of 
% the true positive and false positive candidates for this case is given 
% in Appendix~\ref{tbl:truebuggy}\ref{tbl:falsebuggy}

\begin{figure}[t]
\setlength{\belowcaptionskip}{-8pt}
\includegraphics[width=0.923\linewidth]{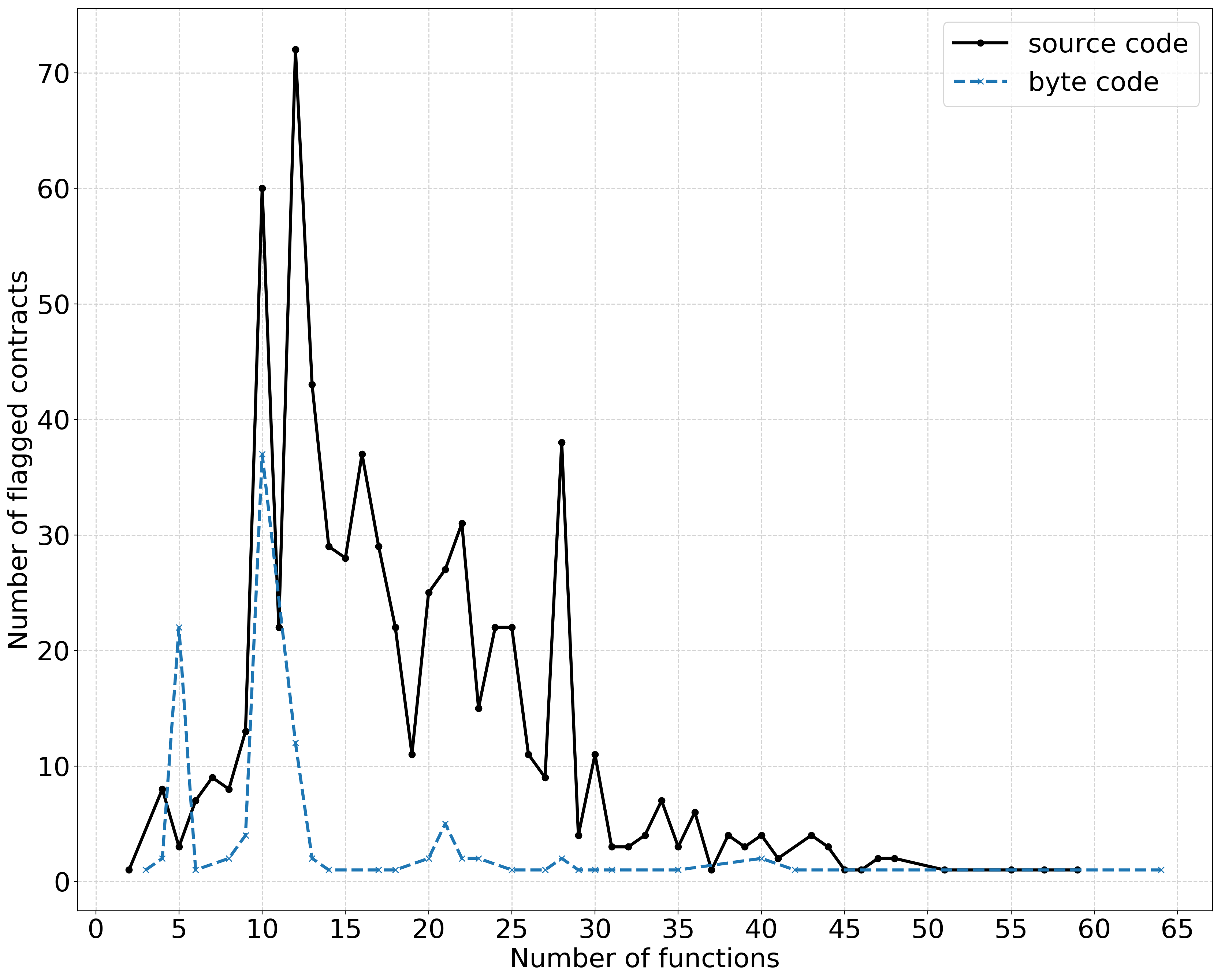}  
\caption{Number of functions in flagged contracts.}
\label{fig:nfunctions} 
\end{figure}

\paragraph{Summary} \toolname produces a small number of concrete traces
that can be analyzed in a few minutes, and more than half of these
correspond to EO violations that we deem to true bugs. \toolname
incurs a practical overhead for diagnosing a dangerous class of errors
before deployment. 

At this point we would like to mention that among all the flagged contracts
with source code for  \Syn violations (i.e., 674), we could find 42 contracts
which  had contacts of their authors mentioned in their source code. We
notified authors of  these contracts and gave a week to take a decision and
reply to us. However, we received no reply  from them in the given time. For
the contracts exhibiting \Lin violations we couldn't find any author to
contact, from reading their source code.

\subsection{Performance}
\label{sec:performance}
About $95\%$ of the contracts require $18.5$ minutes of analysis time
with \tool on average per contract, whereas only $5\%$ of them timeout
after $150$ minutes.  Of the $18.5$ minutes, \tool spends about $15$
minutes to produce events and $3.5$ minutes for fuzzing.

% The efficacy of our $\HB$ extraction in reducing the search space is % evident
% in Figure~\ref{fig:trace_sol}. 
%

Figure~\ref{fig:nfunctions} provides a quantitative explanation for
this efficiency. It shows the total number of functions in contracts;
one can see that a large number of contracts have over $10$ callable
functions, thus analyzing all permutations would lead to prohibitively
large number of traces to test concretely.

A core idea in \toolname is that of filtering out function pairs
before finding their corresponding event pairs in weak \HB-relations.
This technique removes $65\%$ of all pairs on average in our analyzed
contracts. The highest filtration is achieved in a contract with $138$
functions where the tool filters out all but $2$ pairs. The lowest
filtration rate is in $11$ contracts where \tool does not filter out
any of the $4$ found functions. Apart from $11$ contracts out of
$10,000$, \tool provides measurable efficiency gain over the baseline
strategy of enumerating all pairs and checking them symbolically.

\paragraph {An illustrative example} 
\tool finds the \erc bug mentioned in Section~\ref{sec:motivation} in
$5$ minutes, producing a single minimized pair of traces to
inspect. The tool first collects all $11$ functions of the
contract. The pure function analysis marks $8$ functions as pure,
leaving only for $3$ functions for re-ordering. The pure function
analysis substantially improves the power of $\HB$ analysis, since
only $3$ out the $11$ functions needed to be checked for $\Hb$
relations.  Without this optimization, our $\Hb$ analysis
(Section~\ref{subsec:extracthb}) during fuzzing would inspect
$\binom{11}{2}$ or $55$ pairs instead of $3$.
After $4$ minutes of symbolic analysis to recover $\Hb$ relations,
\tool creates $7$ concrete events as input to the fuzzer. In $1$
minute, the fuzzer analyzes traces of length ranging from $2$ to $6$
and it analyzes a total of $2,560$ traces.  This number would be
$8,652$ without our $\HB$ relation analysis, which has over $3x$
improvement to concrete fuzzing in this example.  After fuzzing, \tool
outputs $43$ traces which have bugs and its witness minimization
produces a single pair of traces to inspect. This is precisely the
pair presented in Section~\ref{sec:motivation}. The output produced by \toolname for this contract is given in Appendix~\ref{sect:appendix:output}

% % The larger the
% % pairs we recover in the HB-relation, the lesser are the combinations
% % to fuzz. 

% On average, only $35\%$ of the total possible function pairs
% are $\HB$ pairs. The highest filtration is achieved in a contract with
% $138$ functions where the tool detects only $2$ potential $\HB$ pairs.
% The least filtration rate is in $11$ contracts where \tool detects
% none to be $\HB$ pairs; all of these have $4$ functions each.
% The larger the pairs we recover in the HB-relation, the lesser are the
% combinations to fuzz. 

%\vspace{-10pt}
%\vspace{-5pt}

Figure~\ref{fig:hbvsnohb} shows the correlation between the size of
the $\HB$ relation set and the number of potential traces created
by the fuzzer. As shown in the figure, an increase in the number of
$\HB$ relations leads to a decrease in the number of possible
traces the fuzzer needs to create and check. For instance, on
average of all analyzed contracts, having a single $\HB$ relation
reduces the number of possible traces by nearly $42\%$. Moreover, the 
average number of HB relations produced by \toolname for a contract is 
nearly $5.5$.

\begin{figure}[t]
{\setlength{\belowcaptionskip}{-15pt}{
\includegraphics[width=\linewidth]{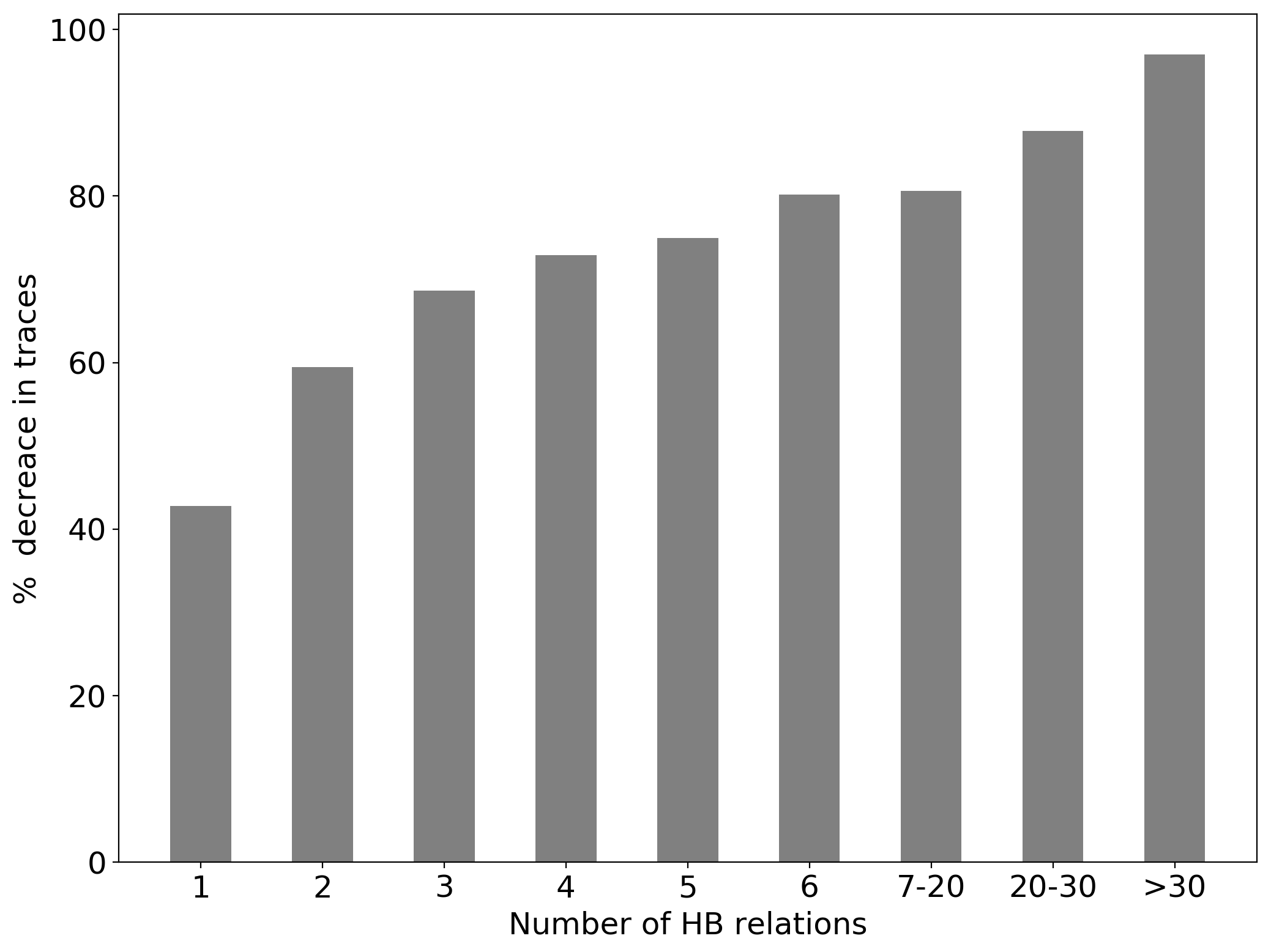}  
\caption{Impact of the number of HB-related event pairs.}
\label{fig:hbvsnohb} 
}}
\end{figure}

\subsection{Comparing \toolname and \oyente}
\label{sec:oy}

\oyente is a symbolic execution analyzer which flags contracts with
two different traces having different Ether flows (\ie, changes in
a contract's balance)~\cite{Luu-al:CCS16,Oyente}. It tries to produce
contexts $m_1$ and $m_2$ such that traces ($e_1, e_2$) and ($e_2,
e_1$) result in two different ether flows where $e_1 = \angled{f_1,
  m_1}$ and $e_2 = \angled{f_2, m_2}$. Here, $f_1$ and $f_2$ can be
the same function. It uses our aforementioned baseline strategy of
enumerating all pairs (traces of depth two) and symbolically
checking if the two paths lead to two different balances.

Our definitions of event ordering (EO) may appear similar to
transaction ordering dependency bugs (TOD) defined and analyzed by the
\oyente tool. There is, however, several definitional and technical
differences between the two tools:

\begin{enumerate}
\item \oyente does not reason about \lin violations; transaction
  ordering dependency bugs are a strict subset of \syn EO violations.

\item \oyente detects two paths with different \code{send}
  instructions, which is unlike EO bugs, characterized by
  observational differences in the final output states. \oyente does
  not account for global state changes other than account balances
  interacting in a transaction. TOD detections by \oyente, thus, may
  include two traces sending to the same recipient, causing no
  modifications to state other than balances.

\item \oyente checks for differences between pairs of traces to
  prevent combinatorial explosion in path space analysis. \tool can
  analyze any combination of traces, and uses a default limit of
  $6$ for presented experiments (but it can be increased).  The key
  reason for its scalability is the use of its HB-relation analysis
  and analysis optimizations.
  % such as (a) prioritizing combinations
  % which do not re-order events of the same function; and (b) filters
  % out functions where the only shared variable is the contract
  % \code{balance}.
  
\item \oyente only reports symbolic path constraints responsible for
  TOD, without giving concrete inputs which exhibit the bug; this is
  unlike \tool which produces concrete and minimized event traces
  for the analyst to inspect. 
\end{enumerate}

\paragraph{Experimental Comparison}
To enable a manual analysis of results, we compared \tool and \oyente
on $5,000$ contracts with available Solidity source code with \oyente
(retrieved on April 27, 2018) and \tool. First, \oyente has no notion
of \lin and thus it detects none of violating contracts that \tool
flags in this category. For \syn violations, \tool flags $674$
contracts, while \oyente flags $251$ cases when its internal default
timeout is set to $150$ minutes (same as \tool) and in online
mode.\footnote{In the online mode, \oyente reads the values of the
  global variables from the actual blockchain, while in offline, it
  assumes they are symbolic variables.} A larger timeout does not
advantage Oyente, since it terminates for all tested contracts in this
timeout.

Out of the $251$ contracts, $78$ are flagged by \tool as well.  We
manually inspect all the remaining $173$ contracts flagged by \oyente.
We find that all these cases are false EO violations --- this confirms
that \tool finds all of true TOD violations flagged by \oyente, and
finds more than double of \syn violations which extend beyond TOD bugs
as well.  We find that \oyente flags TOD false positives mainly due to
two reasons through manual analysis of source code. First, \oyente
assumes that different Ether flows always send Ether to different
addresses, which is often not the case. Second, in contracts with zero
balance, all Ether flows can only send zero Ether, thus they do not
differ in output balances although oyente flags them incorrectly.

%\subsection{Benign and True EO Violations}
%\label{subsec:casestudies}
%

\begin{comment}
\inik{Decide if the below example is good.}
Figure~\ref{fig:oyenteFP} is an example of a false negative in \tool,
found by cross-referencing with \oyente. The missed case is because
\tool optimizes by not fuzzing event histories with the same function
re-ordered. The two events are $e_1 $ $=$ $\angled{\small{\texttt{withdraw}},
  \small{\texttt{owner}}}$ and $e_2$ $=$ $\angled{\small{\texttt{withdraw}},
  \small{\texttt{presaleOwner}}}$; and, two reordered histories are ($e_1$,
$e_2$) and ($e_2$, $e_1$). Line $3$ in the Figure~\ref{fig:oyenteFP}
shows that whoever calls the \code{withdraw} function first gets the
Ether of the contract leaving nothing for the other.  So, in the first
history \code{owner} gets the Ether whereas in second history it
does not.  \toolname does not flag this bug since it does not analyze
the two histories which share the same function \code{withdraw}.

\end{comment}

\section{Related and Future Work}
\label{sec:related}

\vspace{-2pt}

Our work generalizes the class of bugs arising due to
non-deterministic scheduling of concurrent transactions.  The work by
Luu~\etal~\cite{Luu-al:CCS16} identified a specific kind of \syn
violation related to balance updates, and checked for differences
across only a pair of paths. Our work extends this to longer traces
and full state differences, finding nearly $4x$ more true EO
violations of previously unknown bugs.
Our class of bugs are unrelated to liveness or safety properties
identified more recently by symbolic execution
techniques~\cite{Nikolic-al:Maian,Kalra-al:NDSS18,Krupp-Rossow:USENIX18}.
At a technical level, these works provide robust symbolic execution
systems which is equivalent to our assumed starting baseline
technique.  However, in this work, our techniques are complementary to
the baseline, directly addressing the combinatorial blowup due to
checking traces of large lengths. To show this empirically, we
compared directly to a maintained and the open-source \oyente that
operates on EVM bytecode in Section~\ref{sec:oy}. Alternative systems
such as \tname{Zeus}~\cite{Kalra-al:NDSS18} could be used instead of
\oyente, but \tname{Zeus} is not publicly available and operates on Solidity
source rather than EVM bytecode.

It has been later speculated that many of similar issues, stemming
from the non-deterministic transactional nature of smart contracts can
be thought of in terms of shared-memory
concurrency~\cite{Sergey-Hobor:WTSC17}.
For instance, TOD could be considered as a notion of an \emph{event
  race}~\cite{Raychev-al:OOPSLA13,Dimitrov-al:PLDI14}, while DAO and
mishandled responses from \oraclize service would manifest a
violation of logical atomicity~\cite{Herlihy-Wing:TOPLAS90,Grossman-al:POPL18}.
%
% These analogies hinted the possibility of techniques for identifying
% and detecting of a larger class of concurrency-related
% vulnerabilities. 
%
While the knowledge of some of those issues has been accumulated by
the community~\cite{consensys}, we are not aware of any tool that
would implement a principled approach to detect them at scale.

% In our work, we have substantiated the concurrency insights for
% Ethereum smart contracts in a novel way by (a) formulating the notion
% of atomicity/linearizability violation and (b) leveraging the idea of
% intrinsic happens-before ordering induced by a contract implementation
% for efficient detection of non-trivial even traces/TODs, resulting in
% reordering bugs and the corresponding vulnerabilities.
%
% This allowed for automatic discovery of the issues that were out of
% reach for the state-of-the-art tools
% (Section~\ref{sec:motivation}), as well as for efficient
% procedure for detecting of reordering bugs (Section~\ref{sec:tool}).

The most related work to our approach to finding \lin violations
specifically, is the dynamic analysis by
Grossman~\etal~\cite{Grossman-al:POPL18} for detecting of DAO-like
re-entrancy vulnerabilities~\cite{theDao, reentrancy}. Their work
checks for a specific \lin property of \emph{Effectively Callback
  Freeness} (ECF): a contract $c$ is ECF \emph{iff} for any
transaction involving a reentrant call to $c$ one can construct an
equivalent transaction with without reentrant calls.  Their dynamic
analysis employs the notion of \emph{Invocation Order Constraint}
(IOC) graph, which only captures the \emph{fine-grained} shape of a
contract execution within a \emph{single} transaction.  Verifying ECF
dynamically is drastically simpler than challenge tackled by \tool, as
ECF entails checking the commutativity reads/writes of functions under
a \emph{single} transaction execution. In this paper, we are
interested in handling multi-transactional executions and the
associated combinatorial blowup associated. Our notion of
$\HB$-relations and our dynamic symbolic analysis are entirely
different. Consequently, \tool finds $47$ \lin errors and $674$ \syn
errors in about ten thousand contracts, none of which are reported in
the work by Grossman~\etal~\cite{Grossman-al:POPL18}. Their work finds
$9$ contracts which are vulnerable to re-entrancy bugs on the Ethereum
public blockchain, which fall outside the goals and definition of EO
violations.
% 
% This is different from \tool's notion of bugs and
%    that span .  On the
%   other hand, \toolname does not handle reentrancy (which is captured
%   by both \oyente and Grossman~\etal's tools, with the latter
%   reporting 9 vulnerable contracts) due to our adopted execution model
%   (Section~\ref{sec:formal}), in which individual invocations of a
%   contract's methods are considered atomic.
% %
% Instead, \tool targets a substantially larger class of ordering errors
% (violations of serializability and linearizability) across
% multi-transaction traces, as readers can see directly from the new
% examples we presented.  Accordingly, \tool finds 47 distinct
% linearizability-violating contracts, none of which were flagged by
% Grossman~\etal's tool.

A concurrent work reports on the
\tname{Servois}~\cite{Bansal-al:TACAS18} tool for automatically generating commutativity conditions from
data-structure specifications has been successfully used to confirm
the presence of a known \lin violation in a
simple \pname{Oraclize}-using
contract \code{BlockKing}~\cite{Sergey-Hobor:WTSC17}.  However,
\tname{Servois} can only work with an object encoded in a tailored
domain-specific language, thus, it cannot be applied for automatically
detecting bugs at scale of an entire blockchain.

% Several other approaches focused on automated analysis and
% verification of orthogonal behavioral properties of smart contracts,
% which cannot be phrased using the concurrency jargon and require
% different techniques to handle.
%

%
Other approaches based on symbolic analysis of Solidity source code or
EVM implementations have been employed to detect known bugs from a
standard ``smart contract vulnerability
checklist''~\cite{Atzei-al:POST17,quantstamp,Delmolino-al:FC16} in
tools, such as~\tname{Mythrill}~\cite{mythril,mueller-z3},
\tname{SmartCheck}~\cite{smartcheck},
\tname{Securify}~\cite{securify},
and~\tname{Manticore}~\cite{manticore}, none of which addressed EO
bugs.

Our work shares similarity to a vast body of research on checking
linearizability and data race detection in traditional programming
languages~\cite{Bielik-al:OOPSLA15,Flanagan-Freund:PLDI09,Naik-al:PLDI06,Maiya-al:PLDI14,Blackshear-al:OOPSLA18,Burckhardt-al:PLDI10}.
One might expect that some of those tools could be immediately
applicable for the same purpose to smart contracts.
The main obstacle to do so is that, unlike existing well-studied
concurrency models (sequential consistency~\cite{Lamport:CN78},
Android~\cite{Bielik-al:OOPSLA15}, \etc), Ethereum contracts do not
come with the formally specified model of explicit \syn primitives or
have explicit programming abstractions. Because of this, our procedure
for inferring intrinsic $\HB$ relations is considerably different from
prior works. We believe our approach lays out useful abstractions for
future works investigating concurrency in smart contracts.

\subsubsection*{Future Work}
As noted in Section~\ref{sec:formal}, races in a contract can be
manifested not only by execution effects on its local state. Other
vulnerabilities involve non-determinism in choosing a recipient for
transferred funds or logging (aka EVM events). It is possible to
extend \tool's model and implementation to account for those bugs
while fuzzing \wrt constructed $\HB$ relations, and we plan to do so
in the future.

\begin{comment}
% \inik{The below paragraph seems redundant.}
\paragraph{On mechanised verification of smart contracts}
%
While our analysis is designed to detect concurrency-related bugs at
scale, a complementary approach would be to mechanically verify a
contract's implementation to adhere to the desired atomicity/race
freedom properties.
%
At the moment, a number efforts resulted in a complete mechanisation
of EVM semantics~\cite{Gavin-al:yellow-paper} in various frameworks
for interactive and automated proofs:
$\text{F}^{\star}$~\cite{Grishchenko-al:POST18,Bhargavan-al:PLAS16},
Isabelle/HOL~\cite{Hirai:WTSC17,Amani-al:CPP18},
Coq~\cite{Hirai:EVMCoq}, and K~\cite{Hildenbrandt:KEVM17,erc20-k}.
%
It should be possible to encode our properties of interest on top of
those semantics, allowing for the proofs similar to what has been done
before for concurrent
objects~\cite{Dinsdale-Young-al:ESOP17,Sergey-al:PLDI15}, providing
the ultimate safety guarantees.
%
In this regard, we consider our tool to be complementary to those
future efforts, filling the niche modern race detectors occupy for
efficiently finding bugs in deployed concurrent
applications~\cite{Bielik-al:OOPSLA15,Bouajjani-al:POPL15,Flanagan-Freund:PASTE10}.

\end{comment}

% \paragraph{Concurrency in smart contracts}
% %
% Some more related work:
% %
% \begin{itemize}
% \item Adding Concurrency to Smart Contracts~\cite{Dickerson-al:PODC17}
% \end{itemize}

\section{Conclusion}
\label{sec:conclusion}

We have studied event-ordering bugs in Ethereum smart contracts by
exploiting their similarity to two classic notions in concurrent
programs: \lin and \syn violations.  We have provided a formal model
for these violations.  We have shown how to infer intrinsic
happens-before relations from code, and how to use such relations to
shrink the search space when searching for bugs.

%\todo{Novely and impact.} 

\section{Acnowledgements}
\label{sec:ack}

We thank Shweta Shinde and Shiqi Shen for their  valuable comments and their
help with writing a previous version of this paper. We thank sponsors of the
Crystal Center at National University Of Singapore which has supported this
work. Further, Ivica Nikolić is supported by the Ministry of Education, Singapore
under Grant No. R-252-000-560-112. Aquinas Hobor was partially supported by
Yale-NUS College grant R-607-265-322-121. 

%\input{overview}
%\input{issues}
% \input{language}

%% \appendix
%% \section{Appendix Title}
%%
%% This is the text of the appendix, if you need one.
%%
%% \acks
%%
%% Acknowledgments, if needed.

%\setlength{\bibsep}{1.8pt}  
%\newpage

\bibliographystyle{IEEEtranS}
\bibliography{references,proceedings}
\tabularnewline

%% Sorry, no space for appendices!

\appendix
\section{Appendix}
\subsection{Case Studies: True positives}
\label{appendix:casestudies}
In the sequel we present two case study examples of true positives of \syn bugs.

\vspace{5pt}
\paragraph{\textbf{Contest}}
\label{par:contest}
\begin{figure}[t]
\centering
\begin{lstlisting}[mathescape=true,language=Solidity,basicstyle=\scriptsize\ttfamily]
function participate()  {
	participated[msg.sender]=true;
}
function vote(address candidate) {
	if(now < deadlineParticipation || now >= deadlineVoting || voted[msg.sender] || !participated[candidate])
		throw;
	else{
		voters.push(msg.sender);
		voted[msg.sender] = true;
		numVotes[candidate]++;
	}
}
function determineLuckyVoters() {
	for(uint i = 0; i < nLuckyVoters; i++)
		luckyVoters.push(voters[ random(voters.length) ]);
}
function distributePrizes()  {
	for(uint8 j = 0; j < luckyVoters.length; j++)
		luckyVoters[j].send(amount);
	// also sends Ether to the top candidates from numVotes
	...
}
function close(){
	if( now >= deadlineVoting ){
		determineLuckyVoters();
		distributePrizes(); 		
	}
}
...
\end{lstlisting}
%\end{minipage}
%\end{tabular}
\captionof{figure}{A fragment of the contract \code{Contest} }
%\caption{\texttt{Buggy} contract.}
\label{fig:Contest} 
%\end{figure*}
\end{figure}
%\end{multicols}

%
The contract from Figure~\ref{fig:Contest} defines a
Contest\footnote{This is a simplified version of the original contract
  available at
  \url{http://etherscan.io/address/0x325476448021c96c4bf54af304ed502bb7ad0675\#code}.}in
which users can participate and subsequently vote for certain
participants. The voting happens only after participation deadline
(line 5).  Once the voting deadline is met, the contract may be
closed, resulting in distribution of prizes to the participants with
maximal number of votes and to a few lucky voters.

\toolname run on this contract outputs the minimal EO bug
\begin{itemize}
\item
(\code{participate}, \code{vote}, \code{determineluckyVoters}) 
\item  
(\code{participate}, \code{determineluckyVoters}, \code{vote}) 
\end{itemize}

When the function \code{determineLuckyVoters} is called, the
luckyVoters array is updated with a new lucky voter selected
pseudo-randomly from the voters array in line $15$. So, in the first
trace after an account votes for a participant the voters array is
updated with the new accounts' address before calling
\code{determineLuckyVoters}, however, in the second trace, the voters
array is not updated when the \code{determineLuckyVoters} is
called. This leads to different voters array for random selection of
luckyVoters for both traces. Hence, the final list of luckyVoters is
different for both traces.  In order to exploit this bug, the first
voter can always call this function and add him to the lucky voters
array to get the award. In fact, \code{determineLuckyVoters} function
should be called only after the voting deadline for the contest to be
fair for all the voters.

A simple fix to this bug would be to make the determineLuckyVoter
private and only be called inside the close function.

\toolname starts with $29$ functions i.e., $812$ possible pairs of functions
and  narrows it down to $10$ potential hb pairs. Finally, it generates only
$10$ events  in $10 mins$ and feeds them to the fuzzer. In $2 min$ Fuzzer
outputs only $2$ minimal traces of depth $3$ with one of them capturing this
bug. It also outputs a total of $9,592$ traces of depths varying from $3$ to
$6$ which have an EO bug. So, a developer has to analyze only two traces  to
realize the subtle bug.

\paragraph{Escrow Contract}
\label{par:escrow}
\begin{figure}[t]
\centering
\begin{lstlisting}[mathescape=true,language=Solidity,basicstyle=\scriptsize\ttfamily]
contract Escrow  {
  mapping(address => uint) public escrowFee;
  EscrowStruct memory currentEscrow;
  ...
  function setEscrowFee(uint fee) {
  	require (fee >= 1 && fee <= 100);
    escrowFee[msg.sender] = fee;
  }

  function newEscrow(address sellerAddress,
  	             address escrowAddress) payable {
    require(msg.value > 0 && msg.sender != escrowAddress);
    //Store escrow details in memory
    EscrowStruct memory currentEscrow;
    currentEscrow.buyer = msg.sender;
    currentEscrow.seller = sellerAddress;
    currentEscrow.escrow_agent = escrowAddress;
    currentEscrow.amount = msg.value - 
    					escrowFee[escrowAddress] * msg.value / 1000;
    // More manipulations with currentEscrow
    ...
  }
  ...
}
\end{lstlisting}
\caption{A fragment of the contract \scode{Escrow}.}
\label{fig:escrow}
\end{figure}

Consider the fragment of Escrow contract\footnote{The full code of the
  contract is available at
  \url{https://etherscan.io/address/0x1c02ce498dc6d0d6ef05a253e021258b07eeba91\#code}.}
in Figure~\ref{fig:escrow}. The contract facilitate Escrow
transactions between a buyer, a seller, and an escrow agent. A small
fee (called \texttt{EscrowFee}) is associated with every escrow agent
and is paid by a buyer for using the service. The function
\texttt{newEscrow} provides the money storage functionality and
\texttt{setEscrowFee} is used by an escrow agent to set its escrow
fee.

\toolname indicates the minimal EO bug
\begin{itemize}
\item 
(\code{newEscrow}, \code{setEscrowFee}) 
\item
(\code{setEscrowFee}, \code{newEscrow})
\end{itemize}

Notice that both these functions rely on the mapping \code{escrowFee},
by reading and writing to it, which corresponds to the classical
notion of an event race. Generally, an escrow agent would set a fee
before between a buyer and a seller transact. So, \code{setEscrowFee}
has to be called by agent before buyer calls
\code{newEscrow}. However, the contract allows a buyer to call
\code{newEscrow} before any fee is set to the escrow address,
effectively allowing the buyer to forfeit the payment of fee to the
escrow agent.

% \paragraph{Chronoscore}
% \inik{I will update this paragraph only after Aashish settles down on the choice of contract.}
% \label{par:chronos}
% \input{eval-chronos.tex}
% %
% This Contract implements a game which rewards every 7th player who
% joins a game. Every player pays a price to enter a game and the owner
% sets the game parameters such as rewards. \code{play} implements the
% game and \code{setNextGame} is used by the owner to set rewards and
% game timeouts.  \toolname flags two traces (\code{play},
% \code{setNextGame}) and (\code{setNextGame}, \code{play}) and one can
% verify that when a new game round starts it can take the values of
% timeouts and rewards set for the previous round unless owner sets them
% again for the next round. Also, a malicious owner can manipulate the
% game parameters and bias the game in favour of some specific
% address. Similar to the Escrow example above one has to trust the
% owner to not induce any race conditions.
% %
% A simple fix to this bug would be to set the parameters for the next game
% automatically  once the current game ends instead of owner setting them every
% time before a game.

% \input{appendix-graphs}
\subsection{The Output of \toolname on \erc  }
\label{sect:appendix:output}

The following display shows the output of \toolname on one single
buggy instance of the \erc smart contract. Note, this is the list of all pairs of histories that constitute EO bugs (the list of minimized pairs consist of a single pair and it is given in Section~\ref{sec:motivation}.)

\vspace{5pt}

\begin{tiny}

\begin{verbatim}
Events
--------------------------------------------------------------------------------
0 : {'tx_value': '0', 'tx_caller': 'cee827be9b520a485db84d1f09cc0a99ea878686', 'name': 'transfer', 
'tx_input': 'a9059cbb0000000000000000000000006bb8742b27231e22b8f194d4a0737bdeec14c4ca
0000000000000000000000000000000000000000000000000000000000000001'}
1 : {'tx_caller': 'cee827be9b520a485db84d1f09cc0a99ea878686', 'name': 'approve',
'tx_blocknumber': '493e00', 'tx_value': '0', 'tx_timestamp': '5a5c001d', 
'tx_input': '095ea7b3000000000000000000000000cee827be9b520a485db84d1f09cc0a99ea878686
0000000000000000000000000000000000000000000000000000000000000001'}
2 : {'tx_caller': 'cee827be9b520a485db84d1f09cc0a99ea878686', 'name': 'transferFrom',
'tx_blocknumber': '493e00', 'tx_value': '0', 'tx_timestamp': '5a5c001d', 
'tx_input': '23b872dd000000000000000000000000cee827be9b520a485db84d1f09cc0a99ea878686
0000000000000000000000008bad998271c5560b47917b6fd819fd20cf03e789
0000000000000000000000000000000000000000000000000000000000000001'}
3 : {'tx_caller': 'cee827be9b520a485db84d1f09cc0a99ea878686', 'name': 'approve', 
'tx_blocknumber': '493e00', 'tx_value': '0', 'tx_timestamp': '5a5c001d', 
'tx_input': '095ea7b3000000000000000000000000cee827be9b520a485db84d1f09cc0a99ea878686
0000000000000000000000000000000000000000000000000000000000000003'}
4 : {'tx_caller': 'cee827be9b520a485db84d1f09cc0a99ea878686', 'name': 'transferFrom', 
'tx_blocknumber': '493e00', 'tx_value': '0', 'tx_timestamp': '5a5c001d', 
'tx_input': '23b872dd000000000000000000000000cee827be9b520a485db84d1f09cc0a99ea878686
0000000000000000000000008bad998271c5560b47917b6fd819fd20cf03e789
0000000000000000000000000000000000000000000000000000000000000003'}
5 : {'tx_caller': 'cee827be9b520a485db84d1f09cc0a99ea878686', 'name': 'transferFrom', 
'tx_blocknumber': '493e00', 'tx_value': '0', 'tx_timestamp': '5a5c001d', 
'tx_input': '23b872dd000000000000000000000000cee827be9b520a485db84d1f09cc0a99ea878686
000000000000000000000000cee827be9b520a485db84d1f09cc0a99ea878686
0000000000000000000000000000000000000000000000000000000000000001'}
6 : {'tx_caller': 'cee827be9b520a485db84d1f09cc0a99ea878686', 'name': 'transferFrom',
'tx_blocknumber': '493e00', 'tx_value': '0', 'tx_timestamp': '5a5c001d', 
'tx_input': '23b872dd000000000000000000000000cee827be9b520a485db84d1f09cc0a99ea878686
000000000000000000000000cee827be9b520a485db84d1f09cc0a99ea878686
0000000000000000000000000000000000000000000000000000000000000003'}
--------------------------------------------------------------------------------

HB Relations
--------------------------------------------------------------------------------
(1, 2), (3, 4), (1, 5), (3, 6)
--------------------------------------------------------------------------------


Full EO pairs: 43
--------------------------------------------------------------------------------
1 3 2   :  approve approve transferFrom 
1 2 3   :  approve transferFrom approve 
------------------------------------------------------------
1 0 3 2   :  approve transfer approve transferFrom 
1 2 0 3   :  approve transferFrom transfer approve 
------------------------------------------------------------
1 0 3 2   :  approve transfer approve transferFrom 
1 2 3 0   :  approve transferFrom approve transfer 
------------------------------------------------------------
1 0 3 2   :  approve transfer approve transferFrom 
0 1 2 3   :  transfer approve transferFrom approve 
------------------------------------------------------------
1 0 3 2   :  approve transfer approve transferFrom 
1 0 2 3   :  approve transfer transferFrom approve 
------------------------------------------------------------
1 2 0 3   :  approve transferFrom transfer approve 
1 3 0 2   :  approve approve transfer transferFrom 
------------------------------------------------------------
1 2 0 3   :  approve transferFrom transfer approve 
1 3 2 0   :  approve approve transferFrom transfer 
------------------------------------------------------------
1 2 0 3   :  approve transferFrom transfer approve 
0 1 3 2   :  transfer approve approve transferFrom 
------------------------------------------------------------
1 3 0 2   :  approve approve transfer transferFrom 
1 2 3 0   :  approve transferFrom approve transfer 
------------------------------------------------------------
1 3 0 2   :  approve approve transfer transferFrom 
0 1 2 3   :  transfer approve transferFrom approve 
------------------------------------------------------------
1 3 0 2   :  approve approve transfer transferFrom 
1 0 2 3   :  approve transfer transferFrom approve 
------------------------------------------------------------
1 2 3 0   :  approve transferFrom approve transfer 
1 3 2 0   :  approve approve transferFrom transfer 
------------------------------------------------------------
1 2 3 0   :  approve transferFrom approve transfer 
0 1 3 2   :  transfer approve approve transferFrom 
------------------------------------------------------------
0 1 2 3   :  transfer approve transferFrom approve 
1 3 2 0   :  approve approve transferFrom transfer 
------------------------------------------------------------
0 1 2 3   :  transfer approve transferFrom approve 
0 1 3 2   :  transfer approve approve transferFrom 
------------------------------------------------------------
1 3 2 0   :  approve approve transferFrom transfer 
1 0 2 3   :  approve transfer transferFrom approve 
------------------------------------------------------------
0 1 3 2   :  transfer approve approve transferFrom 
1 0 2 3   :  approve transfer transferFrom approve 
------------------------------------------------------------
1 3 5 2   :  approve approve transferFrom transferFrom 
1 5 3 2   :  approve transferFrom approve transferFrom 
------------------------------------------------------------
1 3 2 5 0   :  approve approve transferFrom transferFrom transfer 
1 2 3 0 5   :  approve transferFrom approve transfer transferFrom 
------------------------------------------------------------
1 3 2 5 0   :  approve approve transferFrom transferFrom transfer 
0 1 2 3 5   :  transfer approve transferFrom approve transferFrom 
------------------------------------------------------------
1 3 2 5 0   :  approve approve transferFrom transferFrom transfer 
1 2 3 5 0   :  approve transferFrom approve transferFrom transfer 
------------------------------------------------------------
1 3 2 5 0   :  approve approve transferFrom transferFrom transfer 
1 0 2 3 5   :  approve transfer transferFrom approve transferFrom 
------------------------------------------------------------
1 3 2 5 0   :  approve approve transferFrom transferFrom transfer 
1 2 0 3 5   :  approve transferFrom transfer approve transferFrom 
------------------------------------------------------------
1 3 2 0 5   :  approve approve transferFrom transfer transferFrom 
1 2 3 0 5   :  approve transferFrom approve transfer transferFrom 
------------------------------------------------------------
1 3 2 0 5   :  approve approve transferFrom transfer transferFrom 
0 1 2 3 5   :  transfer approve transferFrom approve transferFrom 
------------------------------------------------------------
1 3 2 0 5   :  approve approve transferFrom transfer transferFrom 
1 2 3 5 0   :  approve transferFrom approve transferFrom transfer 
------------------------------------------------------------
1 3 2 0 5   :  approve approve transferFrom transfer transferFrom 
1 0 2 3 5   :  approve transfer transferFrom approve transferFrom 
------------------------------------------------------------
1 3 2 0 5   :  approve approve transferFrom transfer transferFrom 
1 2 0 3 5   :  approve transferFrom transfer approve transferFrom 
------------------------------------------------------------
1 5 3 2 0   :  approve transferFrom approve transferFrom transfer 
1 0 3 5 2   :  approve transfer approve transferFrom transferFrom 
------------------------------------------------------------
1 5 3 2 0   :  approve transferFrom approve transferFrom transfer 
1 3 0 5 2   :  approve approve transfer transferFrom transferFrom 
------------------------------------------------------------
1 5 3 2 0   :  approve transferFrom approve transferFrom transfer 
0 1 3 5 2   :  transfer approve approve transferFrom transferFrom 
------------------------------------------------------------
1 2 3 0 5   :  approve transferFrom approve transfer transferFrom 
0 1 3 2 5   :  transfer approve approve transferFrom transferFrom 
------------------------------------------------------------
1 2 3 0 5   :  approve transferFrom approve transfer transferFrom 
1 0 3 2 5   :  approve transfer approve transferFrom transferFrom 
------------------------------------------------------------
1 2 3 0 5   :  approve transferFrom approve transfer transferFrom 
1 3 0 2 5   :  approve approve transfer transferFrom transferFrom 
------------------------------------------------------------
0 1 2 3 5   :  transfer approve transferFrom approve transferFrom 
0 1 3 2 5   :  transfer approve approve transferFrom transferFrom 
------------------------------------------------------------
0 1 2 3 5   :  transfer approve transferFrom approve transferFrom 
1 0 3 2 5   :  approve transfer approve transferFrom transferFrom 
------------------------------------------------------------
0 1 2 3 5   :  transfer approve transferFrom approve transferFrom 
1 3 0 2 5   :  approve approve transfer transferFrom transferFrom 
------------------------------------------------------------
1 0 3 5 2   :  approve transfer approve transferFrom transferFrom 
1 5 0 3 2   :  approve transferFrom transfer approve transferFrom 
------------------------------------------------------------
1 0 3 5 2   :  approve transfer approve transferFrom transferFrom 
1 0 5 3 2   :  approve transfer transferFrom approve transferFrom 
------------------------------------------------------------
0 1 3 2 5   :  transfer approve approve transferFrom transferFrom 
1 0 2 3 5   :  approve transfer transferFrom approve transferFrom 
------------------------------------------------------------
0 1 3 2 5   :  transfer approve approve transferFrom transferFrom 
1 2 0 3 5   :  approve transferFrom transfer approve transferFrom 
------------------------------------------------------------
1 3 0 5 2   :  approve approve transfer transferFrom transferFrom 
1 5 0 3 2   :  approve transferFrom transfer approve transferFrom 
------------------------------------------------------------
1 3 0 5 2   :  approve approve transfer transferFrom transferFrom 
1 0 5 3 2   :  approve transfer transferFrom approve transferFrom 
------------------------------------------------------------


\end{verbatim}

\end{tiny}

\end{document}